\documentclass{aa}
\usepackage{hyperref}
\usepackage[varg]{txfonts}

\usepackage{lscape}
\usepackage{multirow}
\usepackage{threeparttable}
\usepackage{array}
\usepackage{booktabs}

\usepackage[normalem]{ulem}

\usepackage{natbib}

\graphicspath{{./}{Figures/}}

\begin{document}

\newcommand{\bysqcm}{cm$^{-2}$}
\newcommand{\Kkmbys}{K km s$^{-1}$}
\newcommand{\bys}{s$^{-1}$}

\newcommand{\glycol}{CH$_2$OHCHO}
\newcommand{\ethanol}{C$_2$H$_5$OH}

\title{Glycolaldehyde and ethanol toward the L1157 outflow: resolved images and constraints on glycolaldehyde formation}

\author{J. Robuschi \inst{1}
            \and  A. López-Sepulcre \inst{1,2}
            \and  C. Ceccarelli \inst{1}
            \and  L. Chahine\inst{1}
            \and  C. Codella \inst{3}
            \and  L. Podio \inst{3}
        \fnmsep\thanks{This work is based on observations carried out under project number I17AB with the IRAM NOEMA Interferometer. IRAM is supported by INSU/CNRS (France), MPG (Germany) and IGN (Spain).}
            }      
            
\institute{Univ. Grenoble Alpes, CNRS, IPAG, 38000 Grenoble, France
\and Institut de Radioastronomie Millimétrique, 38406 Saint-Martin d’Hères, France
\and INAF, Osservatorio Astrofisico di Arcetri, Largo E. Fermi 5, 50125 Firenze, Italy}
  
\titlerunning{Glycolaldehyde and ethanol toward the L1157 outflow}
\authorrunning{Robuschi et al.}

\abstract
{Interstellar complex organic molecules, also known as iCOMs, are species of special interest in astrochemistry because of their potential role in the emergence of life.
Their discovery in the interstellar medium has sparked a decades-long debate about how they are formed.
In principle, two main routes are possible: on the surface of dust grains and in the gas phase.
A powerful way to discriminate between and constrain the two routes is to observe iCOMs along protostellar outflow shocked regions, provided their ages are well-constrained. 
This way, the chemical evolution with time can be probed.}
{In this work, we focus on glycolaldehyde (\glycol) and ethanol (\ethanol), and their possible daughter-mother relationship, which has been suggested by previous works.
More precisely, we aim to verify if the gas-phase reactions derived in these works dominate the formation of glycolaldehyde, thus if this gas-phase formation pathway can account for the abundance of glycolaldehyde we observe in star-forming regions.
We target the well-known southern outflow of L1157, which hosts three shocked regions, B0, B1 and B2, of increasing ages: about 900, 1500 and 2300 yr, respectively.
}
{We obtained high-resolution ($\sim4"$) IRAM NOEMA maps of three lines of glycolaldehyde and one line of ethanol toward the entire southern outflow lobe of L1157 and used them to derive the abundance of the two species in B0, B1 and B2, and their abundance ratio.
We then used a pseudo time-dependent astrochemical code to model the post-shock gas-phase chemistry of the two molecules under study, where glycolaldehyde is formed via gas-phase reactions starting from ethanol, via the so-called ethanol-tree scheme, or on the grain-surfaces.
On the contrary, ethanol is assumed to be formed on the grain-surfaces and released into the gas-phase by the passage of the shocks. 
Once in the gas, \ethanol\, is gradually consumed to form, among other iCOMs, glycolaldehyde.
Ethanol is mainly destroyed by reactions with OH radical, which in turn is primarily formed by injected (previously frozen) water (H$_2$O). 
}
{We present the first ever spatially resolved maps of ethanol and glycolaldehyde toward the L1157 southern outflow and, more generally, toward solar-like star forming regions. 
From the maps, we computed their column densities in B0, B1 and B2. 
Assuming an excitation temperature of 30 K for both species, we found column densities comprised between 1 and 3 $\times10^{13}$ cm$^{-2}$ for glycolaldehyde and 4 and 6 $\times10^{13}$ for ethanol.
Their relative abundance ratio [\glycol] / [\ethanol], equal to 0.25--0.4, increases between B1 and B2.
The measured abundance ratios in B1 and B2 are relatively well reproduced by the astrochemical model. 
However, our model can not simultaneously reproduce the observations toward B0, and B1 and B2, whether we assume that glycolaldehyde is primarily formed in the gas-phase or on the grain-surfaces. 
This likely indicates that one of the assumptions in our model is wrong: the excitation temperature and grain mantle composition, assumed to be the same in B0, B1 and B2, or the age of B0, or the gas temperature, assumed to be constant after the shock passage.
Nonetheless, our modeling rules out the possibility that all the observed gaseous GA is a grain-surface product.}
{The study of resolved iCOM emission in the direction of protostellar molecular outflows proves to be an efficient way to constrain their formation routes. 
In the future, it will be important to carry out similar studies in other regions than L1157 outflow. In addition, better modeling of chemistry in shocks should be developed, in which physical properties vary with time along with chemistry.}


\maketitle

\section{Introduction} \label{sec:intro}  

Among the more than 330 molecules detected in the interstellar medium (ISM)\footnote{See Cologne Database for Molecular Spectroscopy (CDMS) : \url{https://cdms.astro.uni-koeln.de/classic/predictions/catalog/}.}, those with more than 6 atoms and containing carbon plus at least another heavy atom are of particular interest for their potential impact on the emergence of life on Earth and other planets.
These molecules are called interstellar Complex Organic Molecules, iCOMs \citep{Ceccarelli_2017, Herbst_Van_Dishoeck_2009}. 
Glycolaldehyde (\glycol), hereinafter GA, is a particularly interesting iCOM because of its prebiotic potential, as it is the simplest sugar, that may lead to the formation of more complex sugars such as ribose, a key component of ribonucleic acid (RNA), according to biochemists (\citealp{Weber_1998,Jalbout_2007}).

GA was first detected toward the massive star forming region Sagittarius B2(N) by \cite{Hollis_2000} \citep[see also][]{Hollis_2004, Halfen_2006}. 
Subsequently, \cite{Beltran_2009} reported the detection of GA toward the massive hot core G31.41+0.31, marking the first detection of this molecule outside the galactic center.  
\glycol\, has since been detected in other star forming regions, notably toward the solar-like protostars IRAS 16293-2422 \citep{Jorgensen_2012,Jorgensen_2016}, NGC1333 IRAS2A, IRAS4A, IRAS4B1 and SVS13A \citep{Coutens_2015, Taquet_2015, De_Simone_2017}. 
Finally, \glycol\, was detected in the protostellar shocked region B1 of the L1157 molecular outflow by \cite{Lefloch_2017}, using single-dish observations as part of the IRAM Large Program ASAI \citep[Astrochemical Surveys At IRAM:][]{Lefloch_2018}. 

Despite all these detections in various sources, the route of formation of GA, as those of other iCOMs, remains hotly debated.
In general, two possible iCOM formation pathways are evoked in the literature \citep[e.g.][]{Ceccarelli_Faraday, Ceccarelli2023-PP7}: reactions occurring on the dust grain-surfaces \citep{Garrod_2006, Garrod_2008} or in the gas-phase (e.g. \citealp{Vasyunin_2013, Balucani_2015}). 
Discriminating which of the two routes are at work for different iCOMs is all but an easy job and astronomical observations have been used in the past to provide some answers \citep[e.g.][]{Codella_2017, Codella_2020, lopez2024-l1157, Balucani2024-methoxy}.

Regarding GA, \cite{Lefloch_2017} found a good correlation between the abundances of ethanol (\ethanol) and GA, in their observations of L1157-B1 together with three hot corinos (NGC 1333 IRAS 4A and IRAS 2A -- \citealp{Taquet_2015} -- and IRAS 16293-2422 -- \citealp{Jorgensen_2016, Jaber_2014} --).
Inspired by the Lefloch et al. study, \cite{Skouteris_2018} suggested a scheme to form GA in the gas-phase, via a two steps chain of reactions that start from ethanol, called "ethanol genealogical tree".
Their theoretical predictions, obtained via quantum mechanics (QM) calculations, compare extremely well with the observations of GA and ethanol toward the hot corinos mentioned above.
Specifically, the GA-over-ethanol abundance ratios of these sources are consistent with the branching fraction theoretically derived, which holds only in steady-state conditions (valid for the high density in the hot corinos: \citealp[e.g.,][]{Ceccarelli_1996,Taquet_2014}).

A much stronger test can be obtained by adding the constraint of time.
This can be achieved if the emission of the two species is imaged in different shocked regions of a molecular outflow, created by the passage of different shocks from subsequent ejection events.
This method has proven to be successful in constraining the formation route of formamide (NH$_2$CHO), for example \citep{Codella_2017, Lopez-Sepulcre_2024}.
In this article, we use the same method to bring constraints on the formation route of GA.
To this end, we obtained the first ever resolved images of GA and ethanol line emission toward the molecular outflow emanating from L1157-mms and compared our observations with theoretical time-dependent predictions of their abundances.

The article is organized as follows.
In Sect. \ref{sec:The_source:_L1157_southern_outflow} we provide the overview of the target;
in Sect. \ref{sec:observations} we describe the observations;
in Sect. \ref{sec:Results} we present the results of the observations and we derive the column densities and abundances of GA and ethanol along the outflow;
in Sect. \ref{sec:Chemical_modelling} we describe the used astrochemical model and the obtained predictions;
in Sect. \ref{sec:Results_of_the_gas-phase_modelling} we discuss the implications of the comparison between observations and model predictions;
Sect. \ref{sec:Conclusion} summarizes the most important points of the present work.

\section{The source: L1157 southern outflow}
    \label{sec:The_source:_L1157_southern_outflow}

\begin{figure*}
    \title{Zoom on the three shocked regions.}
    \includegraphics[width = \textwidth]{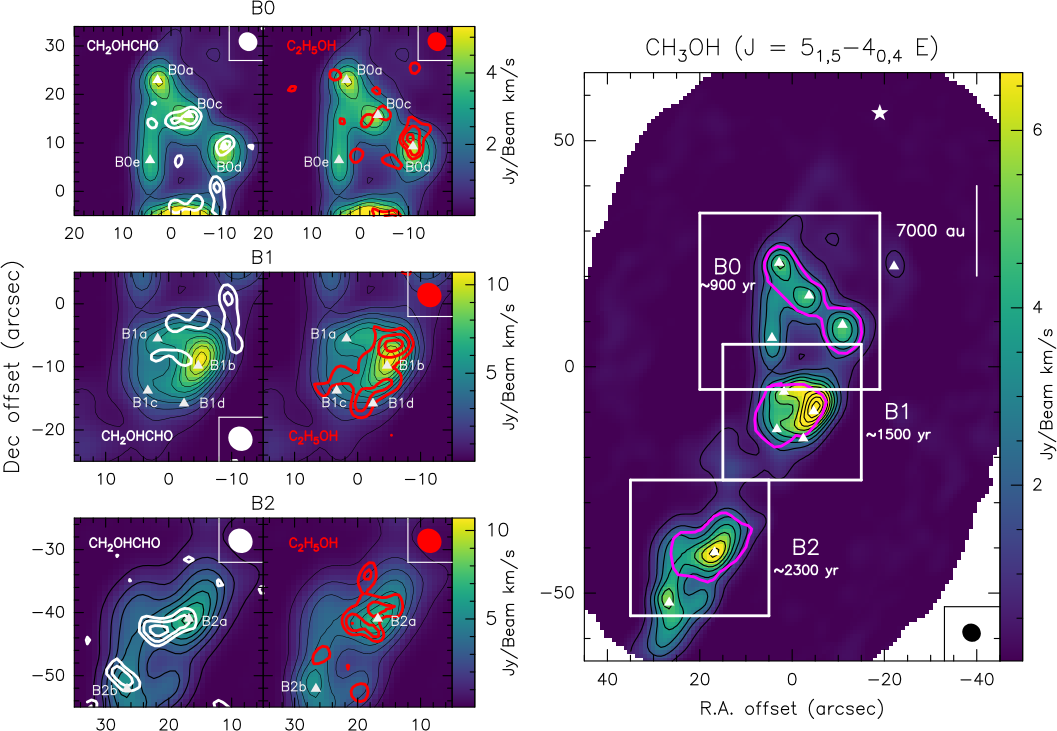}

        \caption{The \textit{left panel} shows a zoom of the GA and ethanol emission maps toward B0, B1 and B2 respectively, integrated over the velocity range $[-10,5]$ km s$^{-1}$. 
        The velocity-integrated emission is shown by white and red contours for the GA and ethanol lines, respectively, overlapped to the emission of the CH$_3$OH line 5$_{1,5}$-4$_{0,4}$ in color scale.
        For the GA and ethanol lines, the first contours are at 3$\sigma$ and the following ones are plotted with a step of 1$\sigma$ (1$\sigma$ = 15 mJy/beam km/s for \glycol\, and 12 mJy/beam km/s for \ethanol\, in B0 and in B1; 18 mJy/beam km/s for \glycol\, and 14 mJy/beam km/s for \ethanol\, in B2).
        The \textit{ellipses} in the \textit{corners} of the maps show the synthesized beams for each GA and ethanol line.
        The \textit{white triangles} and associated labels on all maps mark the different clumps, identified by \cite{Benedettini_2007}. \\
        The \textit{right panel} shows the velocity-integrated intensity map of the CH$_3$OH $5_{1,5}-4_{0,4}$ line, to show the morphology of the overall L1157 southern molecular outflow. 
        The methanol map was used to define the polygons (shown by the magenta contours) where the intensity used for the column density computations is integrated (see text).
        The methanol map is centered at $\alpha(\rm J2000) = 20^{\rm h} 39^{\rm m}09.635^{\rm s}$ and $\delta(\rm J2000) = 68\textrm{°}01'19.80''$. 
        The first contour is at 50$\sigma$ and the next contours are plotted with a step of 100$\sigma$ (1$\sigma$ = 11 mJy/beam km/s). 
        The \textit{black ellipse} in the \textit{bottom right corner} shows the synthesized beam of the CH$_3$OH line imaged. 
        The \textit{white squares} centered on B0, B1 and B2 delimit the respective areas of the zoomed maps presented on the \textit{left panels}. 
        The \textit{white star} at the \textit{top} of map marks the position of the unresolved L1157 protobinary system \citep{Tobin_2013,Tobin_2022}.
        }
     \label{fig:zoom_3_shocks}
\end{figure*}

L1157-mms (\citealp{Tobin_2022}) is a Class 0 protobinary system that drives (at least) an episodic and precessing jet \citep{Gueth_1996, Podio_2016}.
It is is located in the Cepheus molecular cloud system at a distance of 352$\pm$18 pc (\citealp{Zucker_2019}) and has a systemic velocity of +2.6 km s$^{-1}$ (\citealp{Bachiller_1997}).
The L1157 southern outflow lobe, which is blueshifted, extends over a projected length of $\sim$ 0.2 pc (PA$_{\rm outflow} \sim -17$°, see \citealp{Podio_2016}).
This outflow lobe has been the subject of numerous molecular studies (e.g \citealp{Tafalla+Bachiller_1995,Gueth_1996,Arce_2008,Lefloch_2017,Podio_2016,Codella_2017, lopez2024-l1157}), and is known for its richness in iCOMs, making it a perfect target to study their formation. 
The associated jet, whose velocity has been estimated to be $\sim$ 90 km s$^{-1}$ in the southern region, created three shocked regions toward the south, B0, B1 and B2, each of them fragmented into several clumps (\citealp{Benedettini_2007}, see Fig. \ref{fig:zoom_3_shocks}). 

Past studies have revealed that shocked regions are particularly enriched in various molecular species. In particular, strong line emission from SiO and methanol has been detected toward L1157-B1 (\citealp{Mikami_1992, Bachiller_1997}), along with other less abundant species, such as CN, SO, SO$_2$, H$_2$CO, C$_2$H$_5$OH, CH$_3$CHO, HCOOCH$_3$, CH$_3$OCH$_3$, t-HCOOH, NH$_2$CHO, HNCO, CH$_3$CN, etc. (see \citealp{Bachiller_1997, Codella_2009, Mendoza_2014, Lefloch_2017, Lopez-Sepulcre_2024}), whose abundances are enhanced by up to a few orders of magnitude compared with respect to quiescent molecular gas. 
This is explained by the drastic change in the physical conditions due to the shock itself, which causes species that were previously trapped in the icy mantles and refractory cores of dust grains to be released into the gas-phase (\citealp{Bachiller_1995, Bachiller_Perez_Gutierrez_1997}), via sputtering or shattering of the grains. Once in the gas, they can in turn react to form new molecules. 
In addition, B0, B1 and B2 possess different ages, relatively well constrained, and ranging between $\sim$ 900 yr and $\sim$ 2500 yr \citep{Podio_2016}, as summarized in Tab. \ref{tab:Column_densities}.

Put together, this makes L1157's blueshifted outflow an ideal and unique target to test the predictions of astrochemical models in terms of molecular abundance evolution with time.


\begin{table*}
\centering
    \caption{Spectral parameters of the detected molecular lines. All spectral parameters were retrieved from the CDMS database\protect\footnotemark{} \citep{Muller_2016}, which report data from \cite{Pearson_2008} for ethanol, \cite{Widicus_Weaver_2005} for glycolaldehyde and \cite{Xu_2008} for methanol. The latter is imaged to show the outflow morphology (see Sect. \ref{subsec:Maps} and Fig. \ref{fig:zoom_3_shocks}).}
    \label{Table:Spectral_parameters}
\begin{tabular}[h]{cccccccc}
    \hline
    \hline
   Species & Transition & Frequency & $g_{\rm up}$ & $E_{\rm up}$ & $A_{\rm ij}$ & Synthesized beam & PA\\ 
  & & [MHz] & & [K] & [$\times 10^{-5}$\bys] & [$'' \times ''$] & [°] \\
    \hline 
   CH$_2$OHCHO & $8_{0, 8}- 7_{ 1, 7}$ & 82 470.67 & 17 & 18.8 & 1.31 & 4.6 $\times$ 5.0 & $-$225.9  \\
   CH$_2$OHCHO & $8_{1, 8}- 7_{0, 7}$ & 85 782.24 & 17 & 18.9 & 1.49 & 4.7 $\times$ 4.3 & $-$43.2  \\
   CH$_2$OHCHO & $10_{ 1,10}- 9_{ 0, 9}$ & 104 587.74 & 21 & 28.4 & 2.89 & 4.0 $\times$ 3.7 & $-$224.0 \\
 C$_2$H$_5$OH & $7_{0,7} - 6_{1,6}$, anti & 104 487.24 & 15 & 23.3 & 0.79 & 4.0 $\times$ 3.7 & $-$224.0  \\
    CH$_3$OH\tablefootmark{(a)} & 5$_{1,5} - 4_{0,4}$ E & 84 521.17 & 44 & 40.4 & 0.19 & 4.8 $\times$ 4.4 & 137.0 \\
    \hline
\end{tabular}
\end{table*}
\footnotetext{Cologne Database for Molecular Spectroscopy : \url{https://cdms.astro.uni-koeln.de/classic/predictions/catalog/}.}

\begin{table*}
\centering
    \caption{Derived properties of the detected line emission in the different shocked regions.
    The first column lists the shocked regions;
    the second and third columns list the species and observed transitions (Tab. \ref{Table:Spectral_parameters}); the fourth and fifth columns report the measured RMS (measured over a 2 MHz-wide channel) and velocity-integrated (between -10 and +5 km/s) line intensities.
    Note that the upper limits are given at 3$\sigma$.}
\label{tab:Column_densities}    
\begin{tabular}[h]{ccccc}
    \hline
    \hline
  Region & Species & Transition & RMS (1$\sigma$) & $\int_{-10\, \rm{km\, s}^{-1}}^{5\,\rm{km\, s}^{-1}} T_{\rm B} dv$ \\
    \textbf{(\textit{age})}     &         &            & [mK]            & [\Kkmbys] \\
\hline
                       & \glycol      & $8_{0, 8}- 7_{1, 7}$ & 1.6 & 0.11 $\pm$ 0.03\\
                      B0 & \glycol & $8_{1, 8}- 7_{0, 7}$ & 1.7 & < 0.05 \\
                     \textbf{\textit{900 yr}}  &  \glycol       & $10_{1,10}-9_{0, 9}$ & 2.2 & 0.15 $\pm$ 0.04\\
                 & \ethanol & $7_{0,7} - 6_{1,6}$, anti & 2.2 & 0.18 $\pm$ 0.04 \\
\hline
                       &  \glycol & $8_{0, 8}- 7_{1, 7}$ & 2.6 & 0.12 $\pm$ 0.04\\  
                     B1 & \glycol & $8_{1, 8}- 7_{0, 7}$ & 2.3 & < 0.07\\
                      \textbf{\textit{1500 yr}} & \glycol & $10_{1,10}-9_{0, 9}$ & 2.4 & 0.13 $\pm$ 0.04\\ 
                 & \ethanol & $7_{0,7} - 6_{1,6}$, anti & 2.5 & 0.21 $\pm$ 0.04 \\
\hline
                         & \glycol & $8_{0, 8}- 7_{1, 7}$ & 2.9 & 0.18 $\pm$ 0.05\\ 
                      B2 & \glycol & $8_{1, 8}- 7_{0, 7}$ & 2.6 & 0.14 $\pm$ 0.04\\
                      \textbf{\textit{2300 yr}} & \glycol   & $10_{1,10}-9_{0, 9}$ & 2.8 & 0.24 $\pm$ 0.05\\
                 & \ethanol & $7_{0,7} - 6_{1,6}$, anti & 2.8 & 0.21 $\pm$ 0.05\\
\hline
\end{tabular}
\end{table*}



\section{Observations} \label{sec:observations}

\noindent The L1157 protobinary southern outflow was observed at 3 mm with the IRAM NOrthern Extended Millimeter Array (NOEMA) in December 2017, January 2018 and May 2018; for a total of 21 hours in configuration D (the most compact one), using 9 antennas. These observations were made using a mosaic of 8 fields, covering a total area of approximately 2.5 arcsec$^2$. The shortest and longest projected baselines were 24.0 m and 176.0 m respectively, allowing us to recover emission at scales from $\sim$ 3.5$''$ up to $\sim$ 26$''$ at 3 mm. The smallest angular scale is equivalent to $\sim 1230$ au on the source. The phase center is $\alpha(\rm{J2000})=20^{\rm h}39^{\rm m}09^{\rm s}.635$, $\delta(J2000)=+68\textrm{\textdegree}01'19''.80$. We used the PolyFix correlator, covering the frequency ranges 81.5 -- 89.2 GHz and 96.9 -- 104.6 GHz. The molecular lines analyzed in this work were observed with a spectral resolution of 2 MHz (corresponding to $\sim$ 7 km s$^{-1}$ at 85 GHz).
\\
\indent The system temperature varied between 80 and 300 K during most of the tracks, and the amount of precipitable water vapour was comprised between 2.5 mm and 5.5 mm. 3C454.3, 3C84, 3C273, 1928+738 and 2013+370 were used as bandpass calibrators, and the absolute flux scale was fixed by observing MWC349 and LKHA101. The calibration of the gains in phase and amplitude was made on 2010+723, 1928+738 and J1933+656. The estimated uncertainty on the calibrated absolute flux scale is $\leq $ 10\%.

The data were calibrated using the CLIC software from the GILDAS package\footnote{\url{https://www.iram.fr/IRAMFR/GILDAS/}}. We also used the MAPPING software to perform imaging and deconvolution, with a natural weighting to maximize sensitivity. The continuum emission, coming exclusively from the protobinary L1157-mms, was subtracted in the visibility plane. The synthesized beam sizes and 1$\sigma$ RMS achieved are summarized in Tables \ref{Table:Spectral_parameters} and \ref{tab:Column_densities} respectively.

\section{Results}
    \label{sec:Results}

We detected one ethanol and three GA lines, whose $E_{\rm up}$ are comprised between $\sim$ 18 K and 28 K. 
We carefully checked that there was no possible blending with another species for any of the four lines. For example, we excluded the GA transition at $f = 102\, 549.78$ MHz, due to possible line blending with one methyl acetylene (CH$_3$CCH) transition.
The detection of the four lines confirms the previous ones obtained with the 30m IRAM single-dish telescope toward B1 by \cite{Lefloch_2017}. 
The spectral parameters of the detected lines are reported in Table \ref{Table:Spectral_parameters}.

We imaged the emission of the four detected lines and, additionally, one methanol (CH$_3$OH) line (transition $5_{1,5}-4_{0,4}\, E$) with an $E_{\rm up} = 40.4$ K and an Einstein coefficient $A_{\rm ij}=1.9\times 10^{-6}$ s$^{-1}$ (see Table \ref{Table:Spectral_parameters}), for visualization and analysis purposes (see below).

\subsection{Maps and Integrated line intensities}
    \label{subsec:Maps}

The left panels of Figure \ref{fig:zoom_3_shocks} present the GA and ethanol maps toward B0, B1 and B2, respectively, while the right panel shows the methanol map, which provides the overall outflow morphology.
While GA and ethanol have already been detected toward B1 with the IRAM 30-m telescope (\citealp{Lefloch_2017}), this is the first time a resolved map of these two species is obtained in a solar-type star forming region. 

The GA transition $10_{ 1,10}- 9_{ 0, 9}$, displayed in Fig. \ref{fig:zoom_3_shocks}, is the brightest of the three GA lines identified in our data (S/N $\sim$ 6). 
GA emission is detected in a few clumps toward B0, B1 and B2, with the overall brightest emitting regions in B2.
We emphasize that, as it is a mosaic, there are more noise fluctuations on the edges than in the center of the maps shown, with the consequence that the noise in B2 is larger.
Ethanol emission is also seen in clumps toward B0, but it is more extended toward B1 and B2.
Overall, the ethanol line seems to be roughly equally intense in all three shocked regions. 

We estimated the GA and ethanol line intensity recovered by the interferometer toward B1, using the ASAI survey, obtained with the IRAM 30m single-dish telescope.
We found that we recover 100\% of the line flux for each transition (see Appendix \ref{sec:Comparison_with_ASAI} for more details).

In order to compute the molecular column densities toward B0, B1 and B2, we defined polygons based on the methanol line image, as shown in the magenta contours of Fig. \ref{fig:zoom_3_shocks} right panel.

In the following, we describe in detail the maps at B0, B1 and B2, and the chosen polygons.
Table \ref{tab:Column_densities} summarizes the line intensities integrated over the polygons and velocity. 
We note that for each line, we integrated the flux over the channels where there is emission above 3$\sigma$ RMS and, in order to stay consistent, took the same velocity range for all the lines detected.

\subsubsection{B0}
    \label{subsubsec:B0}

The emission of both ethanol and GA is located in the northern part of B0, roughly around the same clumps: B0c and B0d. 
Their spatial distributions match that of methanol quite well; in particular, the three clumps where \glycol\, and \ethanol\, emission is enhanced are also bright in CH$_3$OH (Fig. \ref{fig:zoom_3_shocks}; B0  a, c and d). 
Therefore, we chose the polygon that encompasses the three spots B0 a, c and d.

In B0, we could only detect two GA transitions with a signal-to-noise ratio (S/N) over 3; the $8_{0, 8}- 7_{ 1, 7}$ and the $10_{ 1,10}- 9_{ 0, 9}$. 


\subsubsection{B1}
    \label{subsubsec:B1}

In B1, GA emission is weak and distributed rather to the north of the region.
Ethanol, on the other hand, is relatively bright and extended  mostly toward the west, just below the northernmost GA emission.
Just like in B0, we could only detect two GA transitions, the $8_{1, 8}- 7_{0, 7}$ one having a S/N $\leq$ 3 in this region. 

Around B1, we defined a polygon that encompasses the brightest methanol emission.
We note that extending the polygon to include the northern GA clump emission would result in a too diluted signal with respect to the noise.

\subsubsection{B2}
    \label{subsubsec:B2}

B2 is the shocked region where GA emission is the brightest out of the entire L1157 southern outflow. 
In this region, we notice a slight spatial segregation between GA and ethanol: ethanol seems to be more prominent in the north of B2a, whereas GA is located slightly toward the south compared with ethanol.

The polygon used in B2 encompasses the brightest area around B2a in the methanol emission. 
We excluded the area around B2b because the noise RMS is significantly higher, as it is located at the edge of the mosaic.

\begin{table*}
    \centering
    \caption{Column densities and abundance ratios.
    First column lists the region and, in parenthesis, the age of the shocked regions, where an error of 250 yr has to be considered;
    second, third and fourth columns report the lower limit to $T_{\rm ex}$, and the GA column density  N(\glycol) and abundance \glycol/H, calculated assuming $T_{\rm ex}$=30 K (see text);
    last column lists the abundance ratio GA/Ethanol obtained assuming the same $T_{\rm ex}$ for the two species and computed for $T_{\rm ex}$=20, 50 and 100 K, respectively.}
    \label{tab:abu_ratios}
    \begin{tabular}{lcccc}
    \hline \hline
       Region (age)  & $T_{\rm ex}$ & N(\glycol) (30 K)     & \glycol/H (30 K)   & GA/Ethanol\\
  ~~~~~~~~~~~~[yr] & [K]      & [10$^{13}$ cm$^{-2}$] & [10$^{-9}$]  & ($T_{\rm ex}$=20, 30, 50, 100)\\
       \hline
        B0 (900)     & > 20     & 1.4$\pm0.2$           & $7.0\pm3.5$  & 0.40$\pm0.10$, 0.32$\pm0.07$  , 0.25$\pm0.06$, 0.18$\pm0.04$\\
        B1 (1500)    & > 11     & 1.2$\pm0.2$           & 6$\pm3$ & 0.29$\pm0.07$, 0.23$\pm0.05$  , 0.18$\pm0.05$, 0.13$\pm0.03$\\
        B2 (2300)    & > 17     & 2.3$\pm0.3$           & 12$\pm6$ & 0.55$\pm0.12$, 0.44$\pm0.09$  , 0.33$\pm0.07$, 0.24$\pm0.05$\\
    \hline
    \end{tabular}
\end{table*}

\subsection{Column densities and molecular abundance ratios}
    \label{subsec:Column_densities_and_molecular_abundances_ratios}

We derived the column densities from the velocity-integrated intensities of GA and ethanol lines\footnote{We verified that the spectral parameters retrieved from CDMS and used to derive the column densities are consistent with those listed in the JPL spectroscopy database (Jet Propulsion Laboratory: \url{https://spec.jpl.nasa.gov/ftp/pub/catalog/catdir.html}).}.
To this end, we assumed that the emission is at local thermal equilibrium (LTE), 
all lines are optically thin\footnote{We verified \textit{a posteriori} that the lines are indeed optically thin, and found $\tau < 10^{-3}$ for the 4 GA and ethanol transitions}
and GA and ethanol lines trace the same gas \cite[as discussed in][]{Lefloch_2017}.
Please note that a non-LTE analysis cannot be carried out for lack of collisional coefficients.

When applying the LTE method we found that the GA $8_{1, 8}- 7_{0, 7}$ line (at 85.782 GHz) intensity is systematically too low to fit with the other two GA lines. 
Since the NOEMA integrated interferometric and ASAI data toward B1 agree very well, this implies that the low $8_{1, 8}- 7_{0, 7}$ line intensity is not due to a problem with the NOEMA observations and that it may be due to a non-LTE effect that we cannot reproduce with our analysis.
For example in B1, assuming a GA column density equal to $(1.3\pm 0.2)\times 10^{13}$ cm$^{-2}$ for a $T_{\rm ex}=30$ K (see below), LTE analysis predicts an integrated intensity of $0.12\pm 0.02$ K km s$^{1}$ for the $8_{1, 8}- 7_{0, 7}$ transition, which is two times larger than the measured one. 
Therefore, in the following, we only use the other two GA lines.

\begin{figure}
    \centering
    \includegraphics[width=\linewidth]{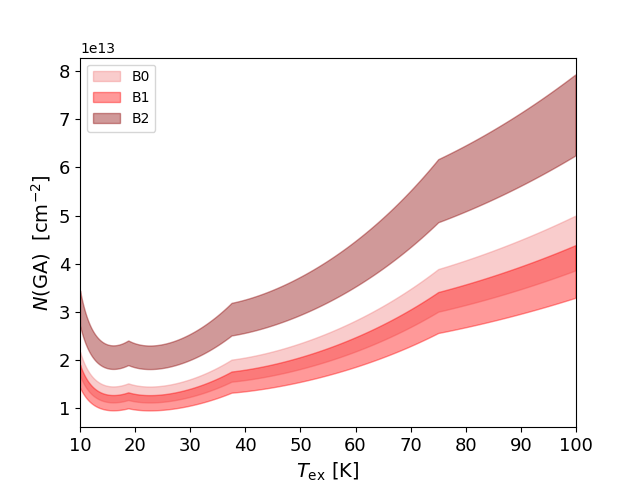}
    \includegraphics[width=\linewidth]{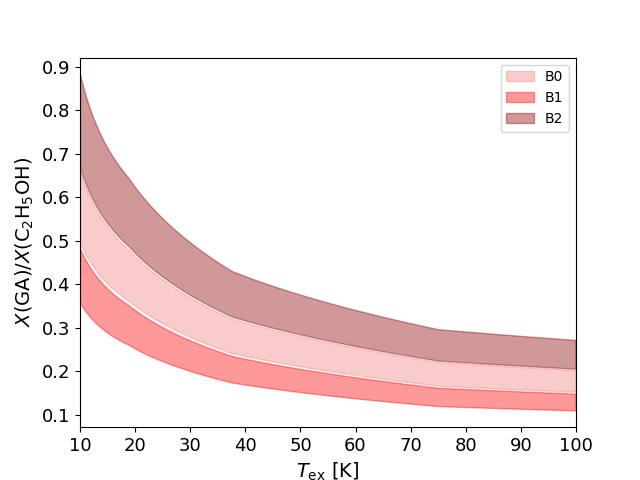}
    \caption{GA column density and abundance ratio, computed from the $10_{1,10}-9_{0, 9}$ line, the brightest one, assuming LTE population and optically thin line emission. The bands represent the uncertainty associated to the line intensity (see text). \textit{Upper panel}: Column density of GA in B0 (light pink), B1 (pink) and B2 (dark pink), respectively, as a function of the excitation temperature $T_{\rm ex}$. Please note that the column density values partially overlap in B0 and B1.
    \textit{Lower panel}: GA/ethanol abundance ratio versus $T_{\rm exc}$ in B0, B1 and B2 (same color coding as above), respectively.} 
    \label{fig:GA_column_density}
\end{figure}

We tried to estimate an excitation temperature $T_{\rm ex}$ with the two GA lines but, given the relatively large error bars on their intensities, we could only derive a lower limit to $T_{\rm ex}$ at each position, as listed in Tab. \ref{tab:abu_ratios}.
Since \cite{Lefloch_2017} found $T_{\rm ex}$=30$\pm$5 K in B1, using 14 GA lines, we computed the column density N$_{\rm GA}$ at 30 K in the three regions, to have a reference column density.
However, we also obtained N$_{\rm GA}$ in the three regions varying $T_{\rm ex}$ from 10 to 100 K, to evaluate the impact of the $T_{\rm ex}$ choice.
The results are shown in Fig. \ref{fig:GA_column_density}.
If we assume the excitation temperature to be the same in the three regions, we observe that N$_{\rm GA}$ is similar in B0 and B1, and significantly larger in B2.
However, it is important to keep in mind that the three temperatures could actually be different, with B0 warmer than B1 and B2 \citep{lopez2024-l1157}.

Using the literature value for the H-nuclei column density in B1, $N_{\rm H}$=$2\times 10^{21}$ cm$^{-2}$ \citep{Lefloch_2012}, we then estimated the GA abundance.
We assume that $N_{\rm H}$ is the same in B0, B1 and B2, based on the CO (J=1-0) emission map by \cite{Gueth_1996}. 
Note that in the literature, $N_{\rm H}$ is reported in the range $(1-4)\times 10^{21}$cm$^{-2}$ for B1 \citep{Bachiller_1997,Lefloch_2012,Mendoza_2014,Codella_2017}, while no estimates exist in B0 and B2.
Thus, we computed the uncertainties on the absolute abundances by taking into account an error on $N_{\rm H}$ of a factor 2. 

Finally, we derived the GA/Ethanol abundance ratio, considering the ratio between the column densities of each species obtained adopting the same $T_{\rm ex}$ and varied it between 10 and 100 K. 
Figure \ref{fig:GA_column_density} shows the dependency of the GA/Ethanol abundance ratio from $T_{\rm ex}$ and Tab. \ref{tab:abu_ratios} lists representative values.
Again, assuming the same temperature in each region, the ratio is larger in B2 than in B0 and B1.
That said, the ratio varies by a factor $\sim$3 at each position across the excitation temperature values considered, i.e. between 10 K and 100 K.
In the following, in order to compare the astrochemical model predictions with observations, we will consider that the three regions (B0, B1 and B2) have the same $T_{\rm ex}$ and we will explore three temperatures: 15, 30 and 50 K.

Note that the errors in the column densities take into account the noise RMS plus a 10\% uncertainty on the absolute intensity scale; 
the errors in the absolute abundances include an additional factor 2 due to the uncertainty of the H-nuclei column densities; 
the abundance ratios do not include neither the calibration nor the $N_{\rm H}$ uncertainty. 

\begin{table*}
\centering
    \caption{Physical parameters and injected species abundances (with respect to H nuclei) used in the chemical modeling (Sec. \ref{sec:Chemical_modelling}). 
    The first column reports the parameters; the second column the values adopted in the Reference model; the third column refers to the articles from which those values are taken; the fourth column gives the range used in the model grid; the last column provides the values of the parameters that best reproduce the observations: please note that there is a degeneracy on the values of temperatures (which was varied between 80 and 130 K) and the abundance of water injected into the gas (between 0.25 and 1 $\times 10^{-4}$. 
    \textit{Upper part:} Physical parameters of the shocked gas: density $n_{\rm H}$, temperature $T_{\rm gas}$ and cosmic-ray ionization rate $\zeta_{\rm CR}$.
    \textit{Bottom part:} Abundances of the species injected into the gas phase after the passage of the shock. 
    Note that we ran two times the Reference model, one assuming no glycolaldehyde on the grain-surfaces and the other assuming that all the observed glycolaldehyde abundance ($7 \times 10^{-9}$) is injected from grain mantles.}
    \label{tab:chemical_modelling_parameters}
\begin{tabular}[h]{c|c|c|c|c}
\hline 
\hline
\multicolumn{5}{c}{Physical parameters of the shocked gas} \\
\hline 
\hline 
Parameter & Reference model & Ref. & Range &Best models (Fig. \ref{fig:Modeling_vs_observations}) \\
\hline 
$n_{\rm H}$ [cm$^{-3}$] & 8.0 $\times$ 10$^{5}$ & 1 & $(4-16)\, \times$ 10$^{5}$ & 8 $\times$ 10$^{5}$\\
Temperature [K] & 90 & 1 & $50-150$ & $80-130$ \\
$\zeta_{\rm CR}$ [s$^{-1}$] & 6.0 $\times$ 10$^{-16}$ & 2,* & $(3.0-12.0) \, \times$ 10$^{-16}$ & 6.0 $\times$ 10$^{-16}$\\
\hline 
\hline
\multicolumn{5}{c}{Injected species abundances, with respect to H} \\
\hline 
\hline 
Species & Reference model & Ref. & Range & Best models (Fig. \ref{fig:Modeling_vs_observations}) \\
\hline 
& & & &  \\
H$_2$O & 1.0 $\times$ 10$^{-4}$ & 3,4 &  $(0.25-4.0)\, \times$ 10$^{-4}$ & $(0.1-1.0)\, \times$ 10$^{-4}$ \\
CO$_2$ & 3.0 $\times$ 10$^{-5}$ & 2,4 & $(1.5-6.0)\, \times$ 10$^{-5}$ & 3.0 $\times$ 10$^{-5}$\\
NH$_3$ & 1.0 $\times$ 10$^{-5}$ & 5 & $(1.0-4.0)\, \times$ 10$^{-5}$ & 1.0 $\times$ 10$^{-5}$\\
CH$_3$OH & 4.0 $\times$ 10$^{-6}$ & 6 & $(2.0-8.0)\, \times$ 10$^{-6}$ & 4.0 $\times$ 10$^{-6}$\\
OCS & 1.0 $\times$ 10$^{-6}$ & 2 & $(1.0-4.0)\, \times$ 10$^{-6}$ & 1.0 $\times$ 10$^{-6}$ \\
SiO & 1.0 $\times$ 10$^{-6}$ & * & $(0.5-2.0)\, \times$ 10$^{-6}$ & 1.0 $\times$ 10$^{-6}$\\
Si &  1.0 $\times$ 10$^{-6}$ & * & $(0.5-2.0)\, \times$ 10$^{-6}$ & 1.0 $\times$ 10$^{-6}$\\
H$_2$CO & 5.0 $\times$ 10$^{-7}$ & 5 & $(0.5-2.0)\, \times$ 10$^{-6}$ & 5.0 $\times$ 10$^{-7}$\\
CH$_3$CH$_2$ & 0 & 5 & $(0-20)$ $\times$ 10$^{-8}$ & 0 \\
CH$_3$CHO & 0 & 5 & $(0-4.0)\, \times$ 10$^{-8}$ & 0 \\
NH$_2$CHO & 0 & 5 & $(0-4.0)\, \times$ 10$^{-9}$ & 0 \\
\hline 
C$_2$H$_5$OH &  2.0 $\times$ 10$^{-7}$  & 7,* & $(5.0-40.0)\, \times$ 10$^{-8}$ &  2.0 $\times$ 10$^{-7}$\\
CH$_2$OHCHO & 0 and 1.5 $\times$ 10$^{-8}$  & * & 0 and 1.5 $\times$ 10$^{-8}$ & 0 \\ 
\hline
\end{tabular}
\\
    \textbf{References:} 1- \cite{Codella_2020}, 2- \cite{Podio_2014}, 3- \cite{Busquet_2014}, 4- \cite{Boogert_2015}, 5- \cite{lopez2024-l1157}, 6- \cite{Benedettini_2013}, 7-\cite{Lefloch_2017}. 
    The * corresponds to abundances not well constrained but used in previous works \citep[see][]{Podio_2014,Codella_2017,Tinacci_2023,Giani_2023, lopez2024-l1157}.
\end{table*}
\section{Chemical modeling}
\label{sec:Chemical_modelling}

We compared the measured abundances with those predicted by the pseudo time-dependent astrochemical model described in \S~\ref{subsec:Model_description}.
Note that, given the uncertainty in the absolute abundance derivation (see discussion in \S~\ref{subsec:Column_densities_and_molecular_abundances_ratios}), we mainly considered the ratio between GA and ethanol abundances, summarized in Table \ref{tab:Column_densities}, rather than the absolute abundances themselves.
In the astrochemical model, we used the most up-to-date chemical reaction network of formation and destruction of GA and ethanol in the gas-phase, summarized in \S~\ref{subsec:chemical_network}.

\subsection{Model description}\label{subsec:Model_description}

We followed the same method as in previous works by our group \citep[e.g.][]{Podio_2014, Codella_2017, Codella_2020, Tinacci_2023, Giani_2023, Lopez-Sepulcre_2024}, and used the GRAINOBLE code (\citealp{Taquet_2012, Ceccarelli_2018}) with the gas-phase only option, and the gas-phase reaction network described in the next section. 

Briefly, the passage of a shock is characterized by a sudden increase of the shocked gas density and temperature, and the partial release into the gas-phase of the components of the dust icy mantles and refractory cores. The latter is due to the sputtering (gas-grain collisions) and shattering (grain-grain collisions) of grains which, depending on the velocity of the shock, can liberate species from the dust icy mantles and fragment their refractory cores.  \citep[e.g.][]{Flower1996-shocksmod, Field1997-shocksmod, Caselli1997-shocksmod, Jimenez2008-shocksmod, Gusdorf_2008}.

Accordingly, in our model, we increase the shocked gas density and temperature as well as the gaseous abundances of species known to be components of the icy dust mantles and refractory grain cores.
In practice, the employed model consists of two stages, simulating the gas conditions before and after the shock passage, as follows:

\noindent
\underline{Stage 1}: 
We derive the steady-state gas-phase abundances of a typical molecular cloud with a temperature $T_{\rm gas}$ = 10 K and H-nuclei density $n_{\rm H}$ = 2$\times$10$^{4}$ cm$^{-3}$. 
They represent the gas chemical composition before the passage of the shock and are used as input of the Stage 2 modeling.

\noindent
\underline{Stage 2}: 
We set the density and temperature of the gas to the values previously derived for the post-shocked gas in L1157-B1, $T_{\rm gas}$ = 90 K and H-nuclei density $n_{\rm H}$ = 8$\times$10$^{5}$ cm$^{-3}$ \citep[e.g.,][]{Lefloch_2012, Codella_2017}.
We also increase the abundances of the species injected into the gas-phase because of the shock passage, again based on previous determination of the abundances as listed in Tab. \ref{tab:chemical_modelling_parameters}.
Finally, we follow the abundance of the various species as a function of time.\\

\subsection{Chemical network: formation and destruction of glycolaldehyde and ethanol}
\label{subsec:chemical_network}

%
\begin{table*}
\centering
    \caption{ 
    Gas-phase chemical reactions involved in the destruction of gaseous ethanol (\textit{upper half of the table}) and in the formation of gaseous OH, a major destroyer of ethanol (\textit{lower half of the table}).
    The first two columns report the reactants and products, the following three columns the $\alpha$, $\beta$ and $\gamma$ parameters.
    In reactions 1, 2 and 5, the rate constant $k$ is computed by the usual Arrhenius-like equation: $k(T) = \alpha ~\left(\frac{T}{300}\right)^{\beta}\exp^{-\gamma/T}$.
    In reaction 3, $\alpha$ represents the branching ratio of the channel and $\beta$ the langevin rate.
    In reaction 4, the following holds: $k(T)=\alpha~\beta ~\left( 0.62+0.4767 ~\gamma~\left(\frac{300}{T} \right )^{0.5} \right)$. 
    Last column lists the references to the works and databases that report the reactions.}
    \label{tab:chemical_reactions}
\begin{tabular}[h]{rclccccc}
\hline
\hline 
Reactants & & Products & $\alpha$ & $\beta$ & $\gamma$ & Label & Ref. \\
         & &           & [cm$^{3}$/s] & & [K] & &\\
\hline
\multicolumn{8}{c}{\textit{Ethanol destruction}}\\
C$_2$H$_5$OH + OH & $\rightarrow$ & CH$_3$CHOH + H$_2$O & 1.90$\times$10$^{-11}$ & $0$ & 0 & 1a & \cite{Skouteris_2018} \\
C$_2$H$_5$OH + OH & $\rightarrow$ & CH$_2$CH$_2$OH + H$_2$O & 8.10$\times$10$^{-12}$ & $0$ & 0 & 1b & \cite{Skouteris_2018}  \\
C$_2$H$_5$OH + H$_3$O$^+$ & $\rightarrow$ & H$_2$O + C$_2$H$_5$OH$_2^+$ & 1.79$\times$10$^{-9}$ & $-0.50$ & 0 & 2 & KIDA, see text\\
\hline 
\multicolumn{8}{c}{\textit{OH formation}}\\
H$_2$O + NH$_3^+$ & $\rightarrow$ & NH$_4^+$ + OH & 1.0 & 9.39$\times$10$^{-10}$ & 5.41      & 3 & KIDA \& UMIST \\
CH$_3$OH$_2^+$ + e$^-$ & $\rightarrow$ & H + CH$_3$ + OH & 4.54$\times$10$^{-7}$ & $-0.59$ & & 4 & UMIST \\
H$_3$O$^+$ + e$^-$ & $\rightarrow$ & H + H + OH & 2.60$\times$10$^{-7}$ & $-0.50$ & 0        & 5 & UMIST \\
\hline


\end{tabular}
\end{table*}
%
We used the reaction network GRETOBAPE-gas described in \cite{Tinacci_2023}, updated according to the more recent works by \cite{Giani_2023}, \cite{Giani2025-HC5N} and \cite{Balucani2024-methoxy}.

In the following, we review the reactions involved in the formation and destruction of GA and ethanol, the two target species of this work.

\paragraph{Glycolaldehyde}
\label{subsubsec:Glycolaldehyde}

Until a few years ago, there were no known GA gas-phase formation routes that could explain its observed abundance in the interstellar medium. 
For instance, \cite{Halfen_2006} proposed that a sequence of gas-phase reactions, triggered by gaseous formaldehyde (H$_2$CO) reacting with its protonated form (H$_2$COH$^+$), could lead to the synthesis of GA. 
However, later modeling by \cite{Woods_2012} proved the route to be inefficient at the ISM temperatures. 
Meanwhile, given the difficulty of finding gas-phase reactions for the formation of several other iCOMs \citep[e.g., see the discussion in][]{Garrod_2006, Ceccarelli2023-PP7}, the astrochemical community focused its attention on the possibility that GA is formed on the cold dust grain-surfaces during the prestellar phase \citep[e.g.][]{Garrod_2006}.
In the case of GA, \cite{Garrod_2008} proposed that it is formed via the association of HCO with CH$_2$OH.
However, later quantum chemistry computations by \cite{Enrique-Romero_2022} showed that the HCO + CH$_2$OH reaction on icy surfaces (as it is the case in the ISM) has a barrier of $\sim1.7$ kJ/mol (equivalent to $\sim200$ K) and, more importantly, it is in competition with the H abstraction from HCO to form CO + CH$_3$OH, a reaction which has no activation barrier making it more probable than the GA formation.
In other words, the formation of GA via this grain-surface route is likely inefficient.

Based on the observed correlation between ethanol and GA \citep{Lefloch_2017}, \cite{Skouteris_2018} proposed an alternative way to form GA in the gas-phase, starting from gaseous ethanol (see below for its formation), that they called "ethanol tree".
Briefly, in this scheme, ethanol reacts with OH (and Cl) to form the CH$_3$CHOH and CH$_2$CH$_2$OH radicals.
In turn, the reaction of both radicals with O leads to the formation of formic acid (HCOOH), acetic acid (CH$_3$COOH) and acetaldehyde (CH$_3$CHO) from CH$_3$CHOH, and formaldehyde (H$_2$CO) and GA from CH$_2$CH$_2$OH.
\cite{Skouteris_2018} also carried out quantum chemistry calculations and provided the rate of formation of all the products and their branching fraction. 
The reaction forming GA shows a weak dependence on the temperature \citep[$\propto T^{0.16}$:][]{Skouteris_2018}.

Finally, gaseous GA is mainly destroyed by reactions with abundant molecular ions (HCO$^+$, H$_3$O$^+$, H$_3^+$).

In the present work, we mainly focus on the gas-phase formation route of GA, namely the "ethanol tree" scheme of \cite{Skouteris_2018}.
Our goal is to verify whether it can reproduce the observed GA abundance or additional routes forming GA on the grain-surfaces are absolutely necessary.
We emphasize that we do not provide any specific constraint on the grain-surface routes for the following reason. 
The possible amount of GA formed on the grain-surfaces would be completely model dependent because our observations are only sensitive on how much GA is injected into the gas-phase, not how it was formed.
That said, we also tested the possibility that GA is entirely due to grain-surface formation and injected into the gas-phase.

\paragraph{Ethanol}\label{subsubsec:Ethanol}

There are no known efficient reactions forming ethanol in the gas-phase, so it is believed to be formed on the dust grain-surfaces, as many saturated molecules such as methanol \citep[e.g.][]{Tielens1982-mantles, HamaWatanabe2013-ChemRev, Rimola2014-methanol}.

\cite{Garrod_2008} proposed that ethanol is formed by the combination of the radicals CH$_2$OH with CH$_3$, similarly to the formation route proposed by the same authors for GA, discussed above.
Again, quantum chemistry calculations by \cite{Enrique-Romero_2022} showed that the CH$_2$OH + CH$_3$ reaction has an activation barrier of 2.5 kJ/mol (=300 K) and it is in competition with the formation of CH$_4$ + H$_2$CO.
Later, \cite{Garrod2022-newModel} proposed that ethanol is formed by the combination of CH$_3$CH$_2$ with O on the grain-surfaces, a reaction in competition with the hydrogenation of the former.
The same year, \cite{Perrero2022-ethanol} proposed that ethanol can be formed by the reaction of CCH, landing from the gas onto the iced grain-surface, with one molecule of the water ice enveloping the dust grains, followed by hydrogenation.
Finally, \cite{Ferrero2024-Creactions} showed that an alternative route for ethanol formation, efficient in the early phases, would be the reaction of C atoms landing on the ice grain-surfaces forming COH$_2$, followed by the reaction with CH$_3$, barrierless, and hydrogenation.
In summary, at least four routes of ethanol formation on the grain-surfaces have been invoked in the literature.

Once injected into the gas-phase by the shock passage, ethanol is destroyed by reactions involving abundant molecular ions, especially H$_3$O$^+$, like in the case of GA above, plus the reaction with OH, which is also the first step toward the formation of GA \citep[see above and][] {Skouteris_2018}.
The ethanol destruction reactions and their rate coefficients are summarized in Tab. \ref{tab:chemical_reactions}.

\noindent
\textit{Reaction (1), C$_2$H$_5$OH + OH:} We adopted the rate constants reported in \cite{Skouteris_2018}, based on laboratory measurements by \cite{Caravan2015-OH+ethanol} in the 293--54 K range.
Note that this reaction is the first step of the "ethanol tree" leading to the GA formation.

\noindent
\textit{Reaction (2), C$_2$H$_5$OH + H$_3$O$^+$:} We adopted the rate constants from the KIDA database \cite[\url{https://kida.astrochem-tools.org/}:][]{Wakelam2024-KIDA}.
We note, however, that the most recent release of the UMIST database reports a factor 1.6 larger rate constants \citep[\url{https://umistdatabase.uk}:][]{Millar2024-UMIST}, based on old laboratory measurements by \cite{Bohme_1979}, later confirmed by \cite{Spanel_1997}.
The dependence of the rate constant as a function of the temperature is assumed to follow the classical Su-Chesnavich model, i.e. $k\propto T^{-0.5}$.
We notice, however, that more recent and sophisticated treatments were developed for the temperature dependence: the adiabatic capture centrifugal sudden approximation (ACCSA) by \cite{Clary_1985} and the statistical adiabatic channel model (SACM) by \cite{Troe_1996}. 
Both treatments predict rate constants as low as factor 10 with respect to the Su-Chesnavich model \citep[see also the discussion in][]{Ascenzi2019-DMEdestruction}. 
Therefore we kept the value from KIDA and verified that varying the rate constant by a factor 10 did not have much impact on the amount of ethanol destroyed.

Table \ref{tab:chemical_reactions} shows that OH is one of the two major ethanol destruction species, H$_3$O$^+$ being the other one.
Given the OH abundance importance in the observed ethanol abundance, which is different from the injected one, we summarize the major reactions forming OH in Tab. \ref{tab:chemical_reactions} and discuss in the following the adopted rate constants.

\noindent
\textit{Reaction (3), H$_2$O + NH$_3^+$:} We adopted the value reported in KIDA, which is the same in UMIST, based on laboratory measurements by \cite{ANICICH1977-ions}.

\noindent
\textit{Reaction (4), CH$_3$OH$_2^+$ + e$^-$:} We adopted the value reported by UMIST database, which is based on laboratory measurements by \cite{Geppert2006-ch3oh2+}. 

\noindent
\textit{Reaction (5), H$_3$O$^+$ + e$^-$:} We adopted the values reported in the UMIST database, which is based on laboratory measurements by \cite{Novotny2010-H3O+recombination}.
Note that KIDA reports a value, $2.6\times 10^{-7}$, 3.7 times larger than that.

\begin{figure}[!h]
    \centering
    \includegraphics[width=.5\textwidth]{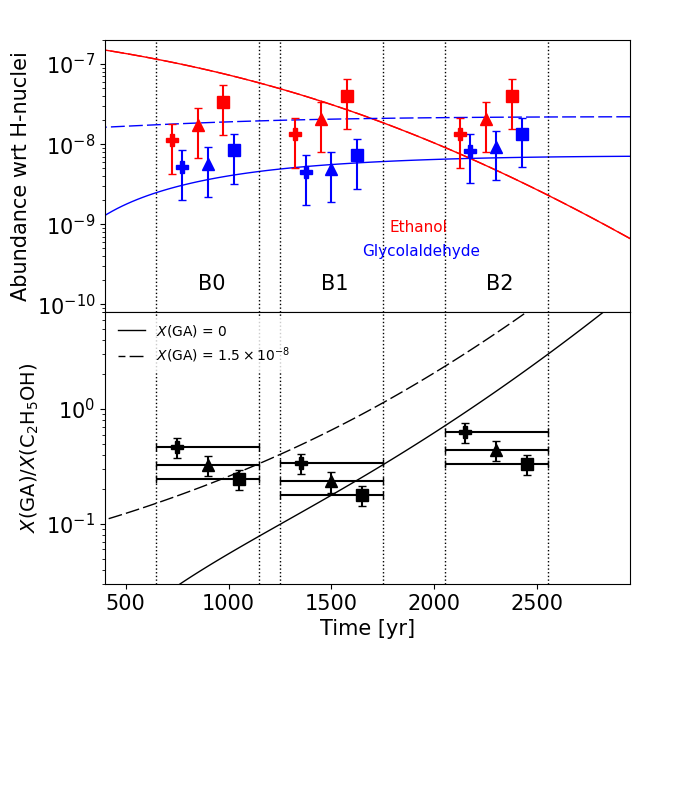}
    \vspace{-2.5cm}
    \caption{Results of the chemical modeling obtained for the Reference Model parameters (listed in Table \ref{tab:chemical_modelling_parameters}) against the observations. 
    The curves shows the predicted abundances (with respect to H-nuclei) of GA (\textit{blue}) and ethanol (\textit{red})  (\textit{upper panel}), and the GA over ethanol abundance ratios (\textit{lower panel}) as a function of time. 
    The symbols represent the values assuming $T_{\rm ex}$ equal to 15 (crosses), 30 (triangles) and 50 (squares) K, as measured at B0, B1 ans B2, respectively.
    The \textit{solid lines} correspond to the Reference Model with no injection of GA from the dust grains, whereas the \textit{dashed lines} correspond to the Reference Model with an injected GA abundance $X(\rm{CH}_2\rm{OHCHO})=1.5\times 10^{-8}$.
    }
    \label{fig:Reference_model}
\end{figure}


\section{Results of the astrochemical modeling}
    \label{sec:Results_of_the_gas-phase_modelling}

In this section, we present the results of the astrochemical modeling described in Sect. \ref{sec:Chemical_modelling} and how they compare with the observations presented in Sects. \ref{sec:observations} and \ref{sec:Results}.

To obtain the models that best reproduce the observations, we first ran a Reference Model with and without the grain-surface contribution to the GA abundance (Sect. \ref{subsec:Reference_model}) and then a grid of models with the input parameters varied according to Tab. \ref{tab:chemical_modelling_parameters} (Sect. \ref{subsec:Best_models}), to understand their impact on the theoretical predictions.
Finally, the comparison between theoretical predictions and observations, as a function of time, allows us to constrain the formations routes of GA. 
We remind that each time corresponds to a position along the outflow.
Indeed, the shocked regions B0, B1 and B2 corresponds to ages of $900\pm250$, $1500\pm250$ and $2300\pm250$ yr, respectively (Tab. \ref{tab:Column_densities}).

\begin{figure*}
    \centering
    \includegraphics[width = .47\textwidth]{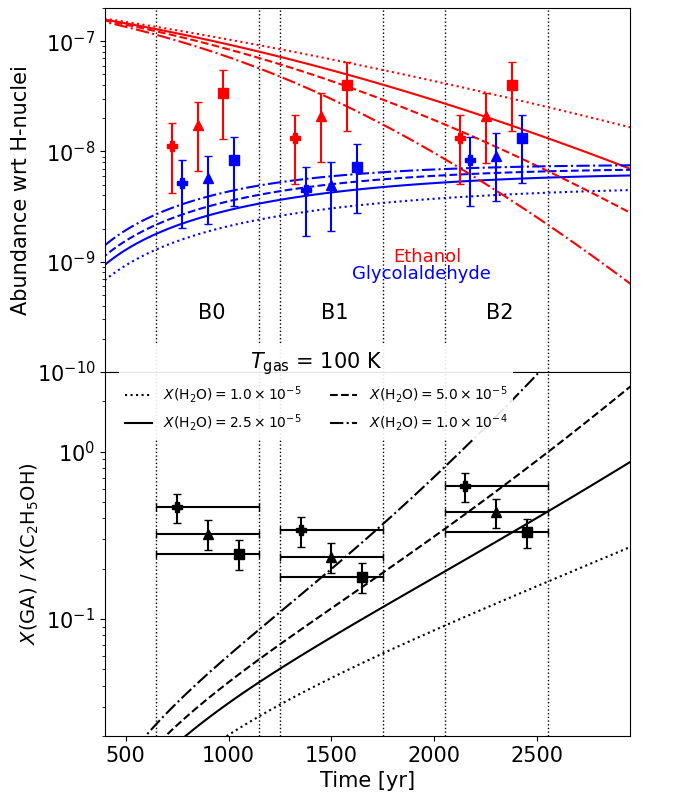}
    \includegraphics[width = .47\textwidth]{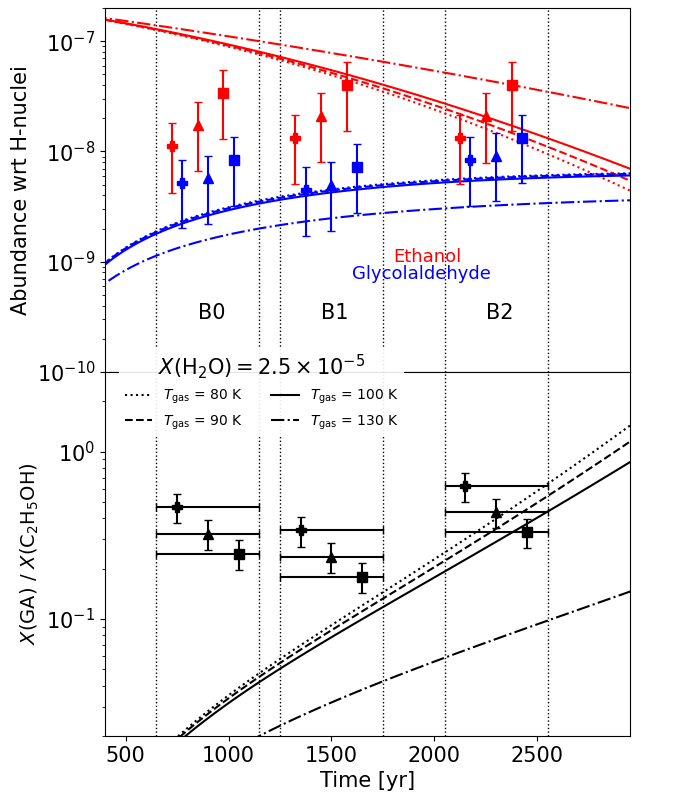}
    \caption{Results obtained assuming that GA is exclusively formed in the gas-phase via the ethanol tree.
    Model predictions (curves) versus observations (symbols) toward B0, B1 and B2, assuming three different excitation temperatures $T_{\rm ex}$: 15 (crosses), 30 (triangles) and 50 (squares) K. 
    In each figure, the \textit{upper panels} show the abundances of GA (blue) and ethanol (red) as a function of time. 
    The \textit{bottom panels} show the evolution with time of the GA/Ethanol abundance ratio.
    \textit{Left panel:} Predictions obtained assuming a temperature of 100 K and four values for the injected H$_2$O abundances: 1.0 (dotted), 2.5 (solid), 5.0 (dashed) and 10 (dotted-dashed) $\times10^{-5}$).
    \textit{Right panel:} Predictions obtained assuming an injected H$_2$O abundance of $2.5\times10^{-5}$ and four values of temperature: 80 (dotted), 90 (dashed), 100 (solid) and 130 K (dotted-dashed).
    }
    \label{fig:Modeling_vs_observations}
\end{figure*}


\subsection{Reference Model}  \label{subsec:Reference_model}

We started the astrochemical modeling from a Reference Model, whose parameters and initial abundances are either retrieved from previous observations or used in past modeling by our group. 
They are listed in Tab. \ref{tab:chemical_modelling_parameters}, along with their references.

Figure \ref{fig:Reference_model} shows the GA and ethanol predicted abundances (with respect to H-nuclei) and their abundance ratio ([\glycol] / [\ethanol]) as a function of time.
The two families of curves show the results by assuming that (i) GA is exclusively formed in the gas-phase via the "ethanol tree" scheme (solid lines) and (ii) GA is mostly formed on the grain-surfaces and injected (at the same time as ethanol) into the gas-phase at the shock passage (dashed lines).
Note that the same quantity of ethanol ($2\times 10^{-7}$) is injected into the gas-phase in both models, such as to reproduce the observed ethanol abundance: therefore, GA is also formed in the gas-phase in model (ii).

\noindent
\textit{Reference Model (i):} 
The model assuming only gas-phase formation for GA predicts an increase of the GA abundance with time (i.e., from B0 to B2), reflecting the chemical timescale of its formation in the gas-phase.
Overall, the model reproduces fairly well the GA absolute abundance in B0, B1 and B2, within the error bars and for each set of assumed $T_{ex}$.
On the other hand, the ethanol abundance is predicted to decrease with time (by about a factor 10 between B0 and B2), due to its efficient destruction by the reaction with OH, the first step in the GA formation reaction scheme (e.g., the ethanol tree). 
However, the ethanol abundance measured does not vary between B0 and B2 by more than a factor 4, assuming the same $T_{ex}$ holds in the three shocked regions.
As a consequence, the GA/Ethanol abundance ratio is not consistent with the observations toward B0, B1 and B2.

\noindent
\textit{Reference Model (ii):} 
Figure \ref{fig:Reference_model} shows the abundance evolution when a relatively large quantity of GA is injected into the gas-phase, $1.5\times 10^{-8}$, to well disentangle it from the contribution of the gas-phase GA production.
At early times ($\leq$1000 yr), the predicted GA abundance is, in this case, larger than that predicted by the gas-phase-only formation model (i) and later flattens to a constant value given by the addition of that directly injected plus that formed by the gas-phase reactions from ethanol.

As for the model (i), the predicted abundance ratios fail to simultaneously reproduce the observed values in B0, B1 and B2, because of the fast destruction of ethanol, as discussed in model (i), and the addition of GA from the grains does not substantially improve the situation. 


\subsection{Grid and best-fit model} \label{subsec:Best_models}

As explained in the previous section, we varied the chemical abundances injected at the passage of the shock, as well as the physical parameters, over the ranges reported in Table \ref{tab:chemical_modelling_parameters}. 
In total we ran about 100 models, where there is not injected GA and, thus, GA is only formed in the gas-phase via the ethanol tree scheme.
Then, we found the best fitting models by minimizing the variance $\chi^2$ defined as: 
\begin{equation}\label{eq:chi2}
\    \chi^2 = \sum_i \frac {\left(theo_i ~-~ obs_i \right)^2}{\sigma_i^2}
\end{equation}
where $theo_i$ and $obs_i$ are the [\glycol] / [\ethanol] abundance ratios, more constraining and reliable than the absolute abundances, respectively predicted at the estimated age of each region B0, B1 and B2, and those measured. 
$\sigma_i$ is the measured uncertainties, which include both the error in the derived abundance ratio $\Delta {obs}$ and the age of the position, as well as the range of the model predicted values within the error of the age $\Delta {theo}$:
\begin{equation}\label{eq:sigma2}
    \sigma^2 = \Delta {obs}^2 ~+~ \Delta {theo}^2
\end{equation}

We found that the models with the set of parameters reported in Tab. \ref{tab:chemical_modelling_parameters}, column 5, best reproduce the observations.
With the exception of the gas temperature, and the injected water and ethanol abundances, all other parameters are fairly well constrained and are the same of the values of the Reference Model (column 2 of Tab. \ref{tab:chemical_modelling_parameters}), confirming the robustness of our previous modeling works \citep[e.g.][]{Podio_2014, Codella_2017, Giani2025-HC5N, lopez2024-l1157}.

\begin{figure*}
    \centering
     \includegraphics[width = .47\textwidth]{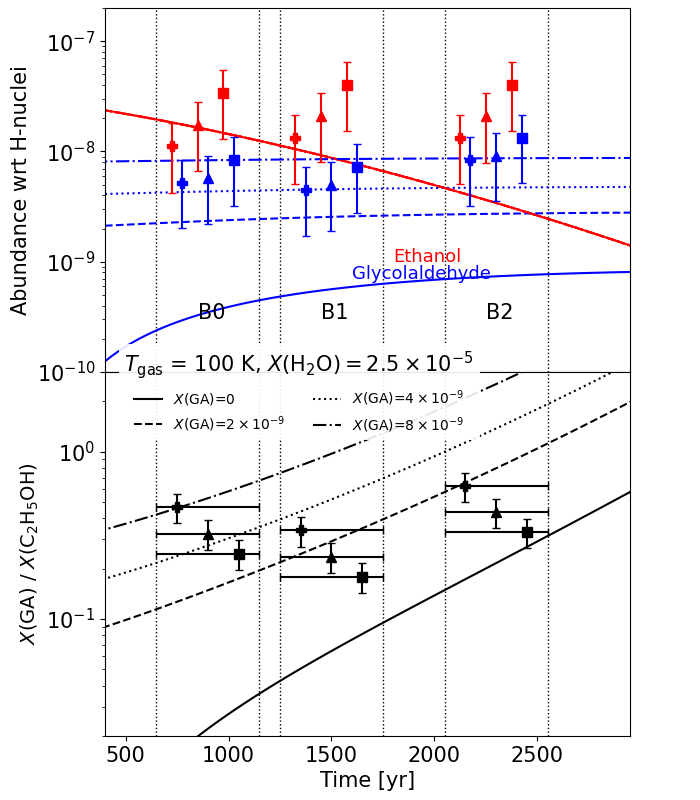}
    \includegraphics[width = .47\textwidth]{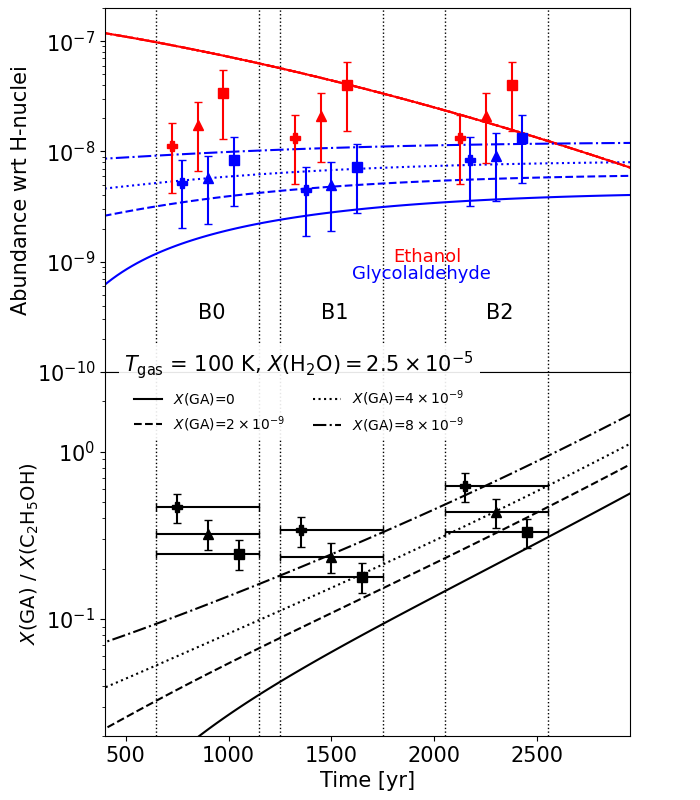}
    \caption{Results obtained assuming that GA is formed in the gas-phase by the ethanol tree and it is injected from the grain mantles into the gas-phase at the passage of the shocks.
    The injected ethanol is 3 (\textit{left panels}) and 15 (\textit{right panels}) $\times 10^{-8}$.
    Model predictions (curves) versus observations (symbols) toward B0, B1 and B2, assuming three different excitation temperatures $T_{\rm ex}$: 15 (crosses), 30 (triangles) and 50 (squares) K. 
    In each figure, the \textit{upper panels} show the abundances of GA (blue) and ethanol (red) as a function of time. 
    The \textit{bottom panels} show the evolution with time of the GA/Ethanol abundance ratio.}
    \label{fig:Modeling_vs_observations_grain_glycol}
\end{figure*}

The gas temperature ($T_{\rm gas}$), the injected water (X(H$_2$O)) and ethanol (X(\ethanol)) abundances are degenerate, having a large impact on the dependency of the ethanol abundance with time and, thus, on the predicted [\glycol] / [\ethanol] abundance ratio.
In particular, predictions of models with a low water abundance, $1.0\times10^{-5}$, better agree with observations for relatively low gas temperatures, $\leq110$ K; intermediate water abundance models, $5.0\times10^{-5}$, need a gas temperature between 100 and 120 K; finally, high water abundance models, $1.0\times10^{-4}$, require larger gas temperatures, between 115 and 130 K (see Fig. \ref{fig:Modeling_vs_observations}, and discussion below).

Modeling of the CO low to high level J observed lines toward B1 have fairly well constrained the temperature, $T_{\rm gas}\sim$60--80 K \citep[e.g.][]{Lefloch_2012}, a value refined to be $\sim 90$ K by further studies \citep[e.g.][]{Codella_2020}.
Therefore, we favor models with a decreased injected water abundance, $2.5\times10^{-5}$,  with respect to the Reference Model and a $T_{\rm gas}\sim 90$ K.
Note that the new value for the H$_2$O injected abundance is consistent with that measured by \cite{Busquet_2014} with the Herschel Space Observatory\footnote{In B1, \cite{Busquet_2014} measured a water column density $N_{\rm{H}_2\rm{O}} = (4.0-10)\times 10^{16}$ cm$^{-2}$, which implies a water abundance $X(\rm{H}_2\rm{O})= (1.0-10)\times 10^{-5}$, assuming $N_{\rm H} \approx (1-4)\times 10^{21}$ \bysqcm.}.
Furthermore, we verified that this change does not affect the latest results that we obtained on the study of formamide in B0, B1 and B2 \citep{Lopez-Sepulcre_2024}.

Figure \ref{fig:Modeling_vs_observations}, upper panels, shows two sets of model predictions against the observations: 
the first set assumes a temperature of 100 K and four values for the injected H$_2$O abundances (1.0, 2.5, 5.0 and 10 $\times10^{-5}$) to show their impact on the predictions.
The second sets assumes an injected H$_2$O abundance of $2.5\times10^{-5}$ and four values of temperature (80, 90, 100 and 130 K) to show its impact on the predicted abundances.
The figure shows that the lower the abundance of injected H$_2$O the slower the destruction of ethanol with time.
Similarly, increasing the temperature has the same effect.
This behavior is due to the formation of OH (the major destroyer of ethanol), whose abundance is proportional to that of water present in the gas-phase, as discussed in Sect. \ref{subsec:chemical_network}.
In addition, the dominant OH formation reactions depend on the inverse of the temperature, i.e. the smaller the temperature the larger the OH abundance\footnote{As discussed in Sect. \ref{subsec:chemical_network}, the rate constant of the ethanol protonation by H$_3$O$^+$ (reaction 2) may be overestimated by up to a factor 10.
On the other hand, decreasing the injected H$_2$O abundance or increasing the gas temperature  decreases the GA/Ethanol abundance ratio because of the lower production of GA from Ethanol.
Please note that lowering the injected ethanol abundance would decrease both the gaseous ethanol and GA: thus, their abundance ratio would remain practically the same.
For completeness, we ran a model with a lowered rate constant for this reaction and verified that the impact is marginal as the other two ethanol destruction routes take over and compensate for it.}.

In conclusion, assuming that (i) GA is entirely produced by the gas-phase reactions started from injected ethanol, and that (ii) the gas temperature and (iii) ice composition are the same in the three regions, we cannot reproduce simultaneously the GA and ethanol observations of B0, B1 and B2.
In other words, one (or more) of the assumptions may be wrong.
In the following, we will discuss each of the three assumptions.

\subsection{GA formation on the grain-surfaces}
In the following, we consider the possibility that GA is formed on the grain-surfaces and injected into the gas-phase at the passage of the shock.
Figure \ref{fig:Modeling_vs_observations_grain_glycol} shows the observations against the model predictions obtained under the assumption that GA is injected into the gas-phase in different quantities.
Please note that the gaseous GA is the sum of the injected GA abundance and the production from ethanol in the gas-phase.

In order to reproduce the range of observed ethanol abundance, we assumed the injected ethanol abundance X(\ethanol) equal to 3.0 and 15 $\times10^{-8}$, and we vary the abundance of the injected GA, from 0 to 8 $\times10^{-9}$.
The figure shows that only the GA abundance is reproduced by the predictions in the three shocked regions simultaneously with an input of GA from the grain mantles, while the GA/Ethanol abundance ratio is not, both assuming a low ($3.0 \times10^{-8}$) and high ($1.5 \times10^{-7}$) ethanol abundance.

In conclusion, assuming that GA is formed on the grain-surface does not simultaneously reproduce the observations toward B0, B1 and B2.

\subsection{Gas temperature and mantle composition in B0, B1 and B2}
\label{subsec:gas_temperature_and_mantle_composition}

To obtain the previous model predictions we assumed the same temperature and the same mantle composition in the three regions B0, B1 and B2.

However, there is no evidence that the gas in the three regions has the same temperature (and density, for what matters).
On the other hand, the discussion above (Fig. \ref{fig:Modeling_vs_observations}) clearly demonstrates that a different gas temperature would lead to a different evolution of the ethanol abundances and, consequently, the GA/Ethanol abundance ratio.
It is reasonable to think that the temperature in B0 is larger than in B2, as also discussed in \cite{lopez2024-l1157}.
The reason is that, after the shock passage, the gas temperature rises (up to thousands K) and then drops (to tens K) \citep{Gusdorf_2008}.
The shock evolution time scale and the amplitude of the temperature peak largely depend on the shock velocity, magnetic field and pre-shock density and the former could range from hundreds to thousands of years before reaching the final temperature.
During this period, ethanol may be destroyed much more slowly than assumed in our relatively low temperature models.
Unfortunately, there are no measurements of the gas temperature in B0 and B2, as mentioned in Sec. \ref{sec:The_source:_L1157_southern_outflow}. Additionally, a major limitation of our modeling is that the physical properties, such as temperature, are kept constant in time. Unfortunately, to this day, we do not have the tools allowing us to vary both the physical conditions of the shock and the gas-phase chemistry using the most up-to-date chemical networks.

Regarding the grain mantle composition, the distance between B0 and B2 is about 5000-7000 au, in principle large enough for it to be relatively different in the two regions.
However, direct observations of the variation of  mantle compositions in low-mass protostars at these scales do not exist.
Evidences based on the deuteration of formaldehyde (H$_2$CO), that changes by more than a factor 2 in the protostar NGC1333 IRAS4A on a scale of 4000 au \citep{Chahine2024-D2CO}, would support the idea of a grain mantle composition (slightly?) different in B0 and B2, but, at this stage, it is impossible to quantify it.
Another reason for a different mantle composition in B0 and B2 is that the former region has already undergone two shocks, the ones that arrived nowadays in B2 and B1, which may have altered its composition.
Finally, the age of B0 may be underestimated, for example because the western part included in the "B0 polygon" belongs to B1 instead.

In summary, in order to conclude whether a different gas temperature and/or grain mantle composition in B0, B1 and B2 are the reason why our model is unable to reproduce the measured GA/Ethanol abundance ratio, one would need more observations (for example to constrain the temperature and density of B1 and B2) and a more sophisticated time-dependent modeling of the shocks.

\section{Conclusions}
\label{sec:Conclusion}

We mapped the GA and ethanol emission along the L1157 southern outflow, with a resolution of $\sim4"$ (equivalent to $\sim 1400$ au), using the IRAM NOEMA interferometer. 
This is the first spatially resolved map of GA and ethanol in a low-mass protostellar region. 
Our map covers three different shocked regions; B0, B1 and B2, whose ages, estimated by previous studies, range from $\sim$900 to $\sim$2300 yr. 
Our conclusions derived from the observations of these three shocked regions are the following: 
\begin{itemize}
    \item We detected three GA lines and one ethanol line with a S/N over five along the L1157 southern outflow, with similar $E_{\rm up}$. 
    While ethanol emission is more or less equally intense in all three shocked regions, that of GA is stronger in B2 than in the other two regions.
    \item Assuming LTE and $T_{\rm ex}$ = 30 K, we computed a GA over ethanol ratio of 0.32$\pm 0.07$ in B0, 0.23$\pm 0.05$ in B1, and 0.44$\pm$0.09 in B2, that the ratio marginally decreases from B0 and B1 and then increases between B1 and B2.
    \item The trend between B1 and B2 is consistent with the gas-phase formation of GA and can be reproduced by our astrochemical model with an injection of ethanol between 1 and 2 $\times 10^{-7}$.
    However, the astrochemical model fails to simultaneously reproduce the observations toward B0, B1 and B2, whether we assume that GA is mostly formed on dust grain-surfaces or in the gas-phase. 
    This suggests either that our assumptions regarding the same excitation temperature and grain mantle composition in the three shocked regions is incorrect, and/or the age of B0 is underestimated, and/or that varying the physical parameters (e.g. gas temperature, density, etc.) with time in the astrochemical model is paramount to reproduce the observations. 
    In the future, it is mandatory to obtain more observed lines of \glycol\, and \ethanol\, to derive the $T_{\rm ex}$ in B0 and in B2.
    \item The GA abundance predicted by our gas-phase astrochemical model is very sensitive to the gas temperature and the amount of water injected. 
    For a $T_{\rm gas}$ consistent with that derived in B1 (i.e. $\sim 100-90$K) by previous works, the H$_2$O injected abundance has to be 1--5 $\times 10^{-5}$. 
    This relatively low water abundance is necessary to slow down the destruction of ethanol in the gas-phase and better agree with the observations.
    \item Since ethanol is present in the gas-phase with an abundance of about 0.4 and 6 $\times 10^{-8}$, GA will necessarily be produced by it, following the gas-phase reactions of the "ethanol tree", and, therefore, the contribution of grain-surface formed GA can not be a dominant source of gaseous GA in the L1157 outflow shocks. 
    \item So far, B1 is by far the most studied shocked region of the L1157 outflow because of its brightness and  richness in iCOMs. 
    However, we have shown that regarding GA, B2 is the brightest region of the outflow. 
    B2, therefore, may be is even richer in iCOMs and is particularly important, especially in the case of products that take time to form in the gas-phase.
\end{itemize}

As in previous studies carried out by our group (\citealp{Podio_2014,Codella_2017,Tinacci_2023,Giani_2023,Lopez-Sepulcre_2024}), this work shows the relevance of using protostellar outflows displaying several shocked regions of different ages in order to probe gas-phase chemical evolution with time. 
Finally, we emphasize that it is important to have self-consistent shock models that take into account the time evolution of the gas temperature and species abundances in the gas-phase simultaneously, and to better constrain the gas temperature and density in B0 and B2.
\\

\textit{Acknowledgments:}
We thank the IRAM-NOEMA staff for their help and support during the observations. 
We warmly thank Antoine Gusdorf for his precious help and advice on the theoretical aspects of the molecular shocks.
We sincerely thank the anonymous referee for careful reading and thoughtful comments.
This project has received funding from the European Research Council (ERC) under the European Union's Horizon 2020 research and innovation program, for the Project “The Dawn of Organic Chemistry” (DOC), grant agreement No 741002.
Linda Podio and Claudio Codella acknowledge the PRIN-MUR 2020 BEYOND-2p (Astrochemistry beyond the second period elements, Prot. 2020AFB3FX), the project ASI-Astrobiologia 2023 MIGLIORA (Modeling Chemical Complexity, F83C23000800005), the INAF-GO 2023 fundings PROTO-SKA (Exploiting ALMA data to study planet forming disks: preparing the advent of SKA, C13C23000770005), the INAF-MiniGrant 2022 “Chemical Origins” (PI: L. Podio), and financial support under the National Recovery and Resilience Plan (NRRP), Mission 4, Component 2, Investment 1.1, Call for tender No. 104 published on 2.2.2022 by the Italian Ministry of University and Research (MUR), funded by the European Union – NextGenerationEU– Project Title 2022JC2Y93 Chemical Origins: linking the fossil composition of the Solar System with the chemistry of protoplanetary disks – CUP J53D23001600006 - Grant Assignment Decree No. 962 adopted on 30.06.2023 by the Italian Ministry of University and Research (MUR). 
This work made use of ASAI “Astrochemical Surveys At IRAM” (Lefloch, Bachiller, Gonzalez et al. 2017) data.


\bibliographystyle{aa}
\bibliography{biblio.bib}


\begin{appendix}

\onecolumn

\section{Glycolaldehyde emission maps along the entire outflow and spectra in the three shocked regions.}
    \label{sec:Glycolaldehyde_maps_spectra}

\begin{figure}[!h]
    \centering
    \includegraphics[width=0.98\linewidth]{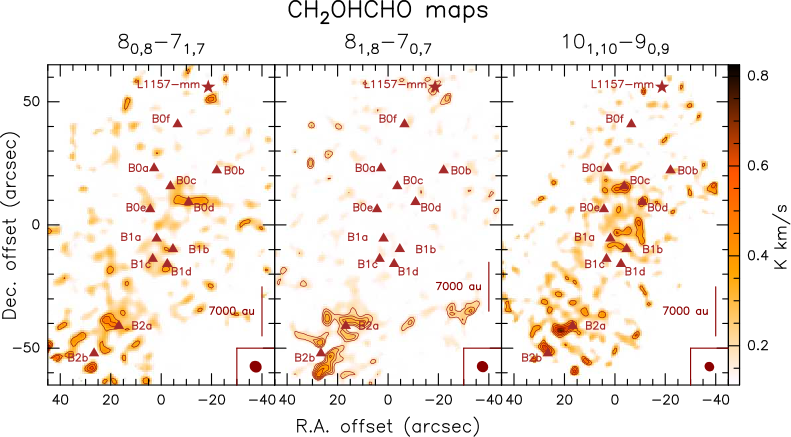}
    \caption{Maps of the emission of GA along the entire southern outflow of L1157, for all three transitions. The first contours are drawn at 3$\sigma$ and the following with a step of 1$\sigma$ (1$\sigma$ = 0.1 K km s$^{-1}$ for the $8_{0,8}-7_{1,7}$ transition, 0.05 K km s$^{-1}$ for the $8_{1,8}-7_{0,7}$ transition and 0.1 K km s$^{-1}$ for the $10_{1,10}-9_{0,9}$ transition). Note that we masked out the noise at the edge of the maps, and that the rms was estimated at the center of the maps; therefore the first contour might be at slightly less than 3$\sigma$ in B2, which is close to the noisier edge of the maps.}
    \label{fig:map-l1157-ch2ohcho-K}
\end{figure}

\begin{figure}[!h]
    \centering
    \includegraphics[width=0.7\linewidth]{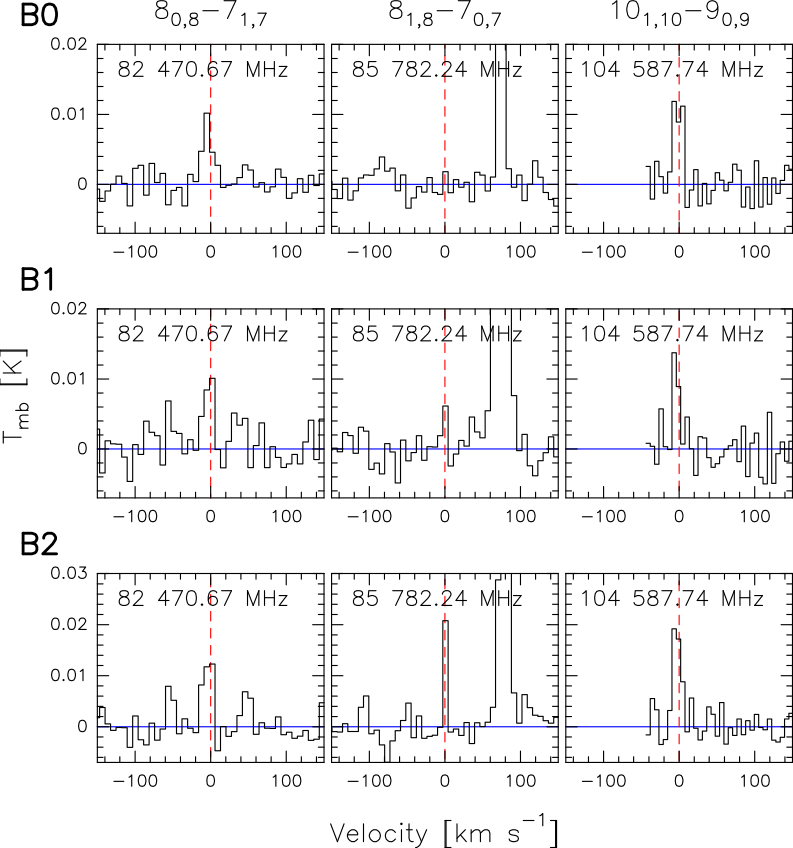}
    \caption{Spectra of the three GA transitions, obtained by integrating the emission inside the polygons defined for each shocked region. The \textit{blue line} marks the zero intensity level of each spectrum, and the \textit{dotted red line} shows the $v_{\rm lsr}$ of L1157-mms, namely $2.6$ km s$^{-1}$ with respect to the Sun.
    Note that the $8_{1,8}-7_{0,7}$ transition is not detected in B0, and is slightly below the 3$\sigma$ threshold in B1.
    Please also note that the emission of the GA transition at $f = 85\, 782.24 MHz$ spans over one channel only, whereas the other two GA lines display emission detected in 3 channels. This is likely due to the low intensity of the former and to the low spectral resolution (spectral resolution of $2$ MHz, equivalent to $\sim 7$ km s$^{-1}$ at $85\, 782$ MHz).}
    \label{fig:ch2ohcho_spectra}
\end{figure}

In this section, we present the emission maps of the three GA lines detected in our data, along the entire southern outflow (Fig. \ref{fig:map-l1157-ch2ohcho-K}). \\
On the whole outflow, all three lines are brighter around B2, where they show clear emission. The $8_{0,8}-7_{1,7}$ and $10_{1,10}-9_{0,9}$ lines also display emission around B0c-d and in B1. Although this emission is fainter than in B2, the two lines are clearly detected in these regions (see spectra integrated over the respective polygons in Fig. \ref{fig:ch2ohcho_spectra}). However, transition $8_{1,8}-7_{0,7}$ is not detected in B0 and marginally detected in B1 (see spectra Fig. \ref{fig:ch2ohcho_spectra}). Note that the emission of the GA $8_{1,8}-7_{0,7}$ transition is also slightly fainter than the other two in the B1 observational data acquired by the large program ASAI (see Sect. \ref{sec:Comparison_with_ASAI} for a comparison with ASAI data). This difference could be due to the low spectral resolution which we are working with (2 MHz) or to the high noise level compared with the line strength.  

\newpage
\section{Comparison with ASAI}
    \label{sec:Comparison_with_ASAI}

\begin{figure}[!h]
    \centering
    \includegraphics[width=.45\textwidth]{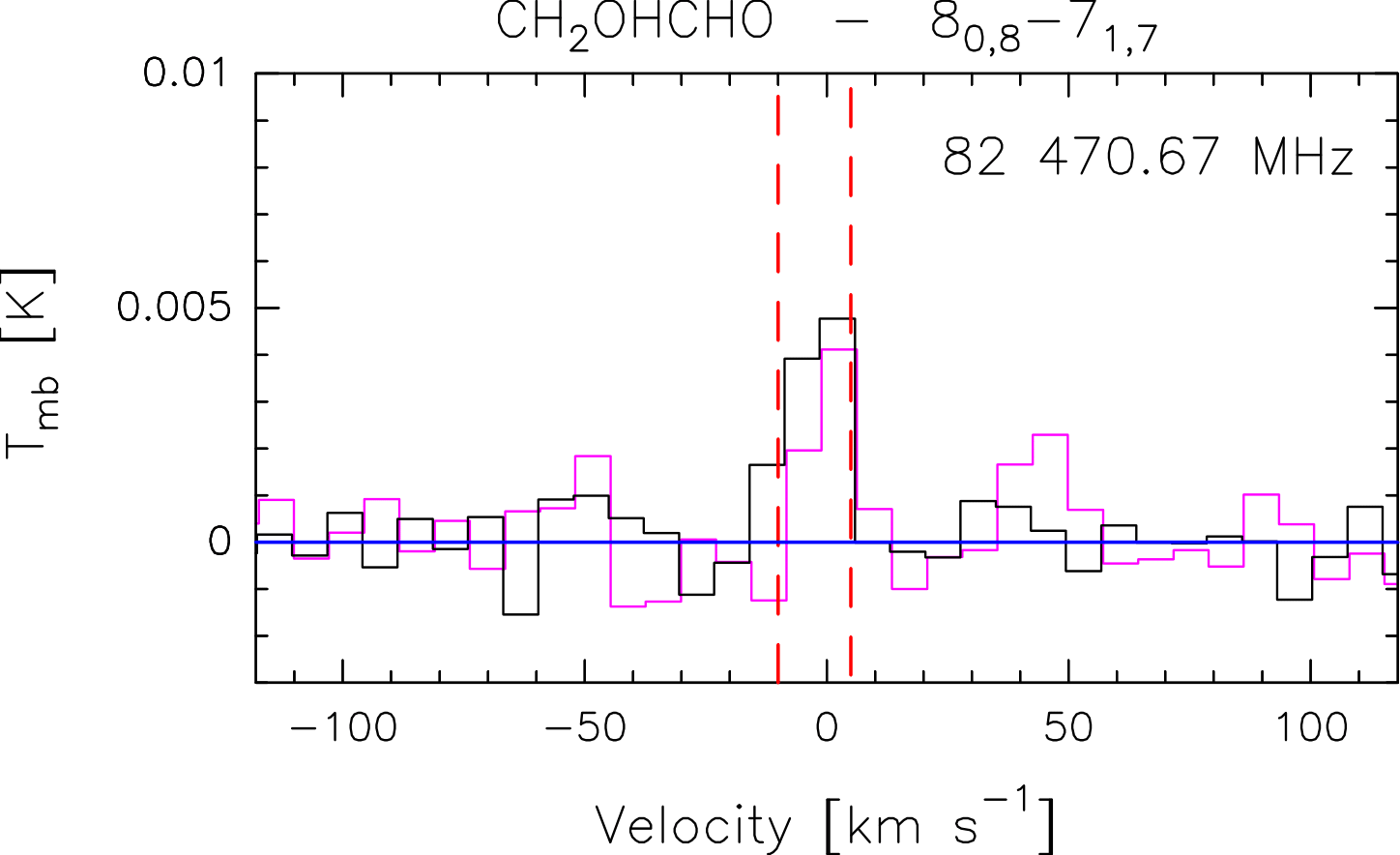}
    \includegraphics[width=.45\textwidth]{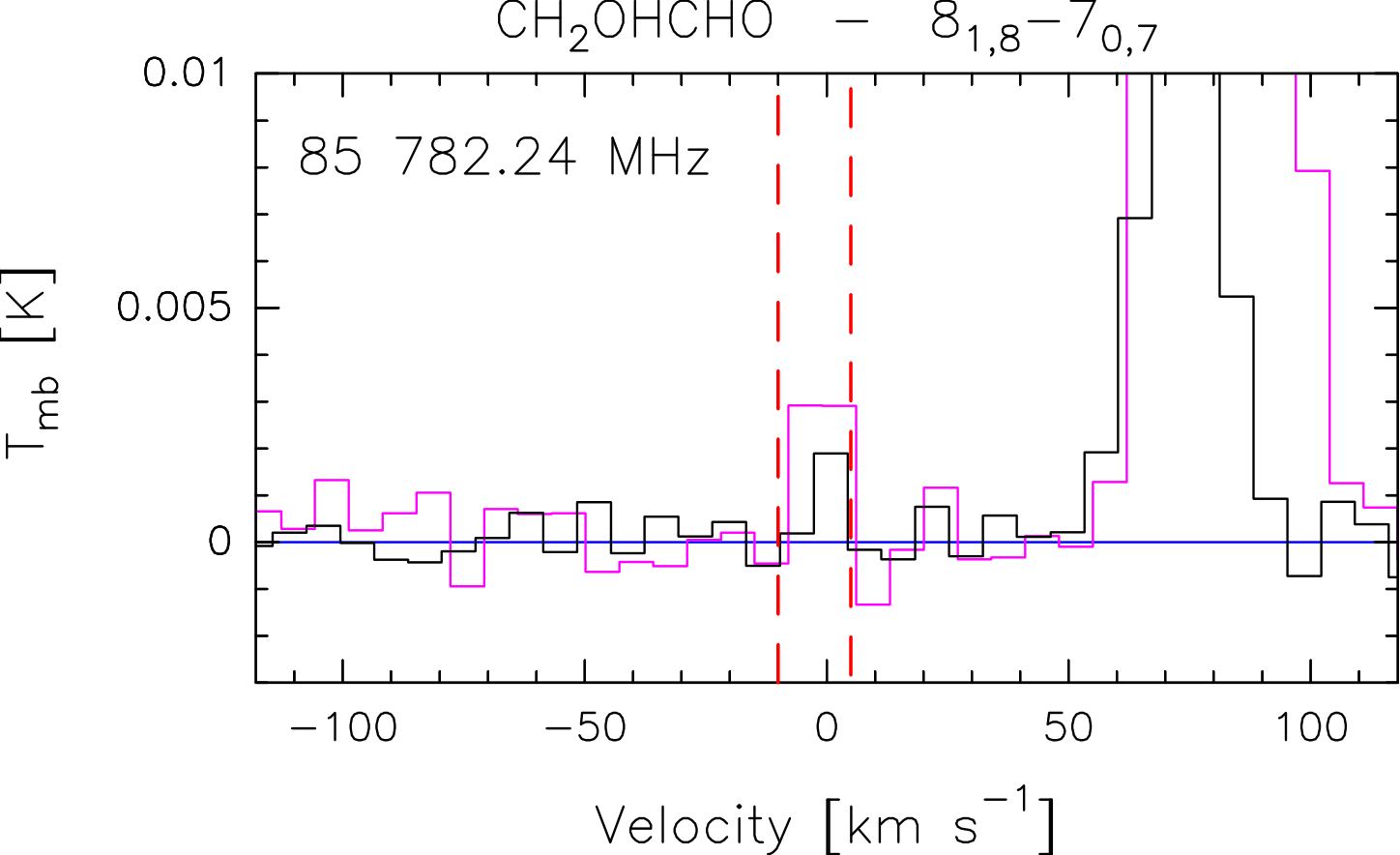}\\~\\
    \includegraphics[width=.45\textwidth]{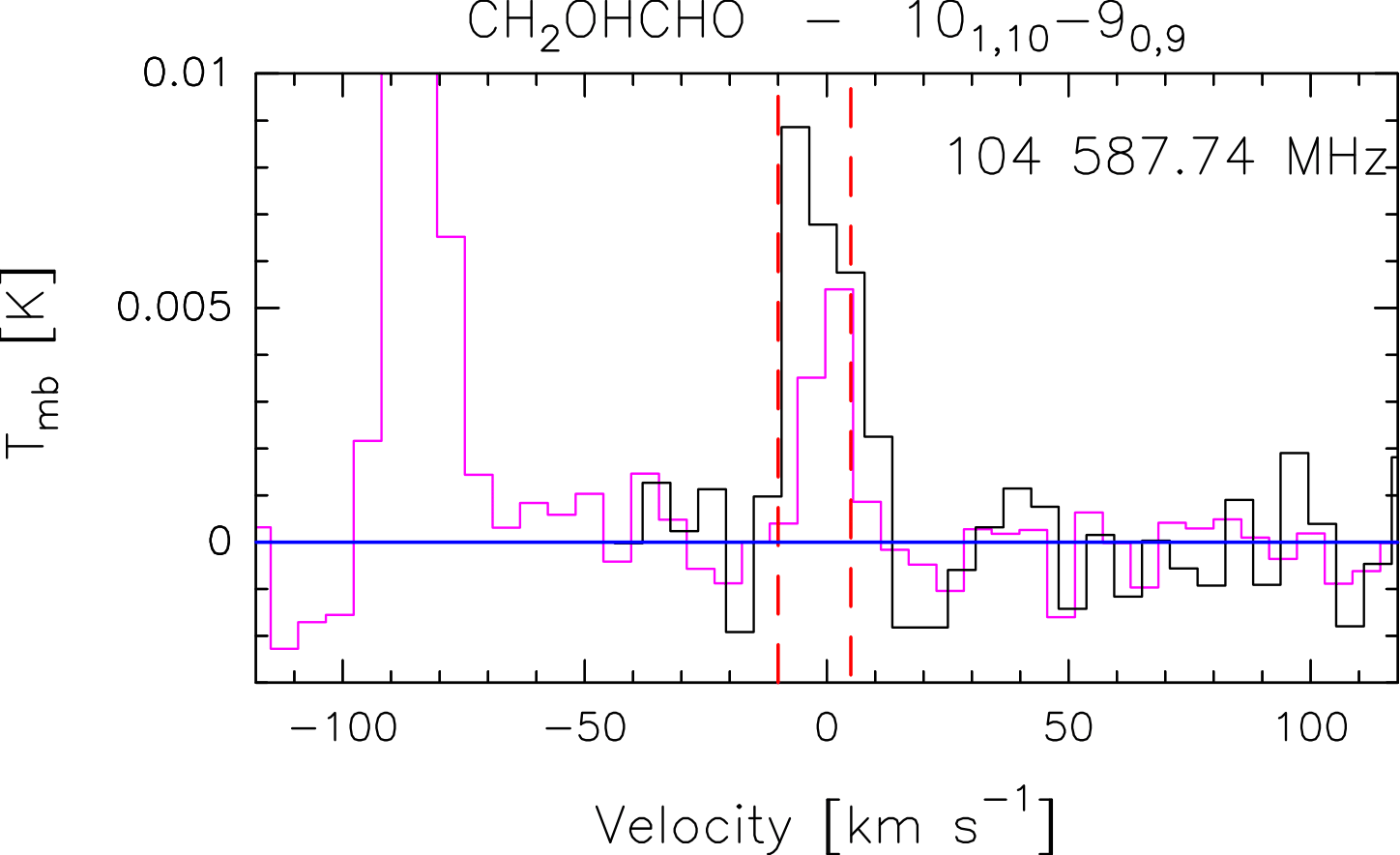}
    \includegraphics[width=.45\textwidth]{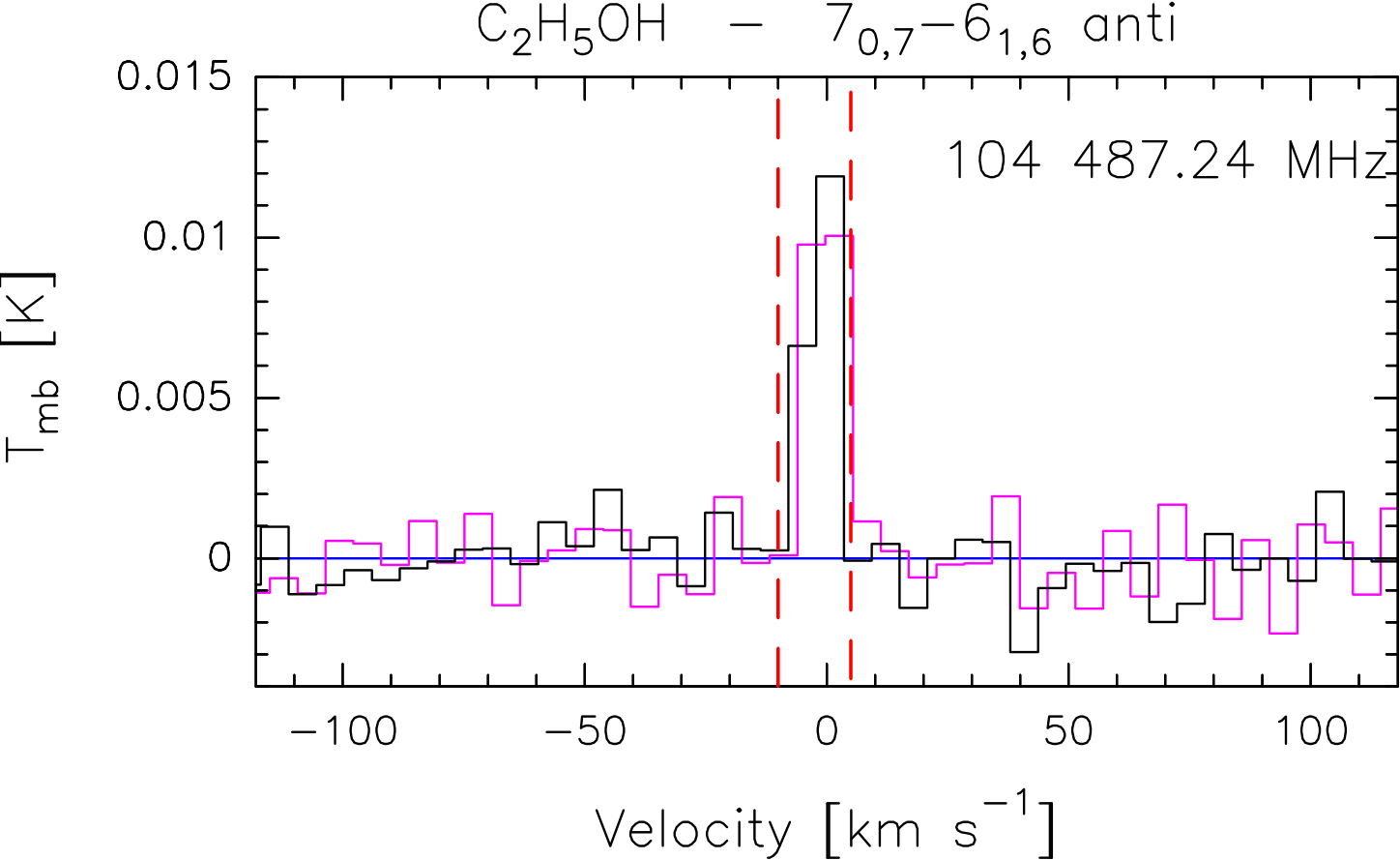}
    \caption{Spectra obtained with the IRAM 30-meter (\textit{in purple}) as part of the ASAI large program vs spectra from our observations obtained with the NOEMA radio-interferometer (\textit{in black}). In order to compare these two datasets, we degraded the spectral resolution of the ASAI data to 2 MHz. The \textit{dotted red lines} mark the velocity range over which the flux was integrated. The NOEMA spectra were obtained by spatially averaging the flux over the half-power beam size of the IRAM 30-m telescope used for the ASAI observations. The \textit{horizontal blue line} marks the zero intensity level of each spectrum.}
    \label{ASAI_vs_NOEMA}
\end{figure}

In order to estimate the fraction of flux filtered out by the interferometer, we compared our results, obtained with the interferometer NOEMA, to those of the large program ASAI (Astrochemical Surveys At IRAM, \citeauthor{Lefloch_2017} \citeyear{Lefloch_2017}, \citeyear{Lefloch_2018}). Indeed, the shocked region B1 located in the southern molecular outflow of L1157-mms has been observed as part of the ASAI large program with the IRAM 30-meter telescope. Using these observations, \cite{Lefloch_2017} detected 14 GA lines and 22 ethanol lines, among which all the lines we detected for these species using NOEMA. 

To conduct this comparison, we averaged the flux recovered around B1 in our data over the IRAM 30-meter telescope half-power beam size, and confronted the spectra with those obtained using the ASAI data\footnote{ASAI data are accessible following this link: \url{https://iram-institute.org/science-portal/proposals/lp/completed/lp007-iram-chemical-survey-of-sun-like-star-forming-regions/}} (see Fig. \ref{ASAI_vs_NOEMA}). We then measured the velocity-integrated intensities for each line in both datasets; the results are reported in Table \ref{tab:Column_densities_ASAI_vs_NOEMA}. In the end, we appear to recover the totality of the ASAI flux with our NOEMA observations. We are consistent with ASAI data within 3$\sigma$, which comforts us in our detection of GA and ethanol. 
We also computed the column densities of ethanol and GA in B1, for our and ASAI observations; the results are summarized in Table \ref{tab:Column_densities_ASAI_vs_NOEMA}. The column densities computed for our GA lines detected in B1, are consistent between the two studies within 3$\sigma$ for an excitation temperature of $\sim$ 30 K (\citealp{Lefloch_2017}).
However, the 3$\sigma$ upper limit on the column density computed using the $8_{1,8} - 7_{0,7}$ GA transition in the NOEMA data is not consistent with the $N_{\rm{CH}_2\rm{OHCHO}}$ computed using the two other transitions of this molecule. Note that this line is also relatively faint in ASAI, falling below their best fit line in their rotational diagram. This could be due to non-LTE effects.

\begin{table}[!h]
\centering
    \caption{Properties of the detected lines - Comparison with those of ASAI.}\label{tab:Column_densities_ASAI_vs_NOEMA} 
\begin{tabular}[h]{cccccc}
    \hline
    \hline
    \\
  Survey & Species & Transition & RMS & $\int_{-10\, \rm{km\, s}^{-1}}^{5\,\rm{km\, s}^{-1}} T_{\rm B}\, dv$ & N 
  \\ 
  & & & [mK] & [\Kkmbys] & [\bysqcm]
  \\~\\
    \hline
  \\

  & CH$_2$OHCHO & $8_{0, 8}- 7_{ 1, 7}$ & 0.9 & $0.04\pm 0.03$ & $(1.4\pm 1.3)\times10^{13}$ \\

  & CH$_2$OHCHO & $8_{1, 8}- 7_{0, 7}$ & 0.6 & $<0.04$ & $< 1.4\times 10^{13}$ \\
 
  ASAI & CH$_2$OHCHO & $10_{ 1,10}- 9_{ 0, 9}$ &  1.1 & $0.05\pm 0.04$ & $(1.1\pm 0.9)\times 10^{13}$ 
  \\~\\
 
   & C$_2$H$_5$OH & $7_{0,7} - 6_{1,6}$, anti & 1.1 & $0.11 \pm 0.04 $ & $(6.5\pm 2.6)\times 10^{13}$ 
  \\~\\
    \hline
  \\
    
  & CH$_2$OHCHO & $8_{0, 8}- 7_{ 1, 7}$ & 0.6 & $0.06\pm 0.01$ & $(2.3 \pm 1.3)\times 10^{13}$ \\ 

  & CH$_2$OHCHO & $8_{1, 8}- 7_{0, 7}$ & 0.6 & $<0.02$ & $<1\times 10^{13}$ \\
   
  NOEMA & CH$_2$OHCHO & $10_{ 1,10}- 9_{ 0, 9}$ & 1.0 & $0.11\pm 0.02$ & $(2.4 \pm 1.2)\times 10^{13}$ 
  \\~\\
  
    & C$_2$H$_5$OH & $7_{0,7} - 6_{1,6}$, anti & 1.1 & $0.11\pm 0.02$ & $(5.5 \pm 3.6)\times 10^{13}$ 
  \\~\\
    \hline
    
\end{tabular}
    \tablefoot{The velocity integrated intensities computed using the NOEMA data and reported in the \textit{5$^{th}$ column} were averaged over the IRAM-30m half-power beam, used in ASAI. 
    The column densities in \textit{column six} were computed using the ASAI (\textit{upper part}) and NOEMA (\textit{bottom part}) data, and considering a beam dilution factor for a source of 20$''$. We derived those quantities assuming LTE and an excitation temperature of 30 K. The uncertainties associated include the 1$\sigma$ noise RMS and a $10\%$ error on the flux calibration.
    Please note that because the flux was integrated over a larger area,  the RMS given in this table differs from the value reported in Tab. \ref{tab:Column_densities}.}
\end{table}

\newpage

\end{appendix}

\end{document}